\documentclass[aps, nofootinbib,superscriptaddress]{revtex4}
\usepackage{amstext,amsmath,amssymb,amsfonts,bbm}
\usepackage[latin1]{inputenc}
\usepackage{fancyhdr}
\usepackage{graphicx}
\usepackage{epsfig}
\usepackage{hyperref}
\usepackage{epstopdf}
\usepackage{pstricks}

\numberwithin{equation}{section}

\def\be{\begin{equation}}
\def\ee{\end{equation}}
\def\bes{\begin{eqnarray}}
\def\ees{\end{eqnarray}}

\newtheorem{definition}{Definition}[section]

\newtheorem{lemma}{Lemma}[section]
\newtheorem{theorem}{Theorem}[section]

\newcommand{\bea}{\begin{eqnarray}}
\newcommand{\eea}{\end{eqnarray}}

\newcommand{\la}{\lambda}

\newcommand{\prf}{\noindent{\bf Proof}\ }
\newcommand{\La}{\Lambda}

\newcommand{\cA}{{\cal{A}}}

\newcommand{\cF}{{\cal{F}}}
\newcommand{\cT}{{\cal{T}}}
\newcommand{\cC}{{\cal{C}}}

\newcommand{\su}{\mathrm{SU}(2)}
\newcommand{\cH}{{\cal{H}}}

\newcommand{\qed}{{\hfill $\Box$}}
\newcommand{\resetequ}{\setcounter{equation}{0}}

\def\la{\langle} \def\ra{\rangle}
\def\f{\frac}

\begin{document}
%%%%%%%%%%%%%%%%%%%%%%%%%%%%%%%%%%%%%%%%%%%%%%%%%%%

\title{\large \bf Scaling behaviour of three-dimensional group field theory}

\author{Jacques Magnen}\affiliation{Centre de Physique Th\'eorique, Ecole polytechnique\\ 91128 Palaiseau Cedex, France}
\author{Karim Noui}\affiliation{Laboratoire de Math\'ematiques et de Physique Th\'eorique, Facult\'e des Sciences et Techniques\\ Parc de Grandmont, 37200 Tours, France}
\author{Vincent Rivasseau}\affiliation{Laboratoire de Physique Th\'eorique, Universit\'e Paris XI\\ 91405 Orsay Cedex, France}
\author{Matteo Smerlak}\affiliation{Centre de Physique Th\'eorique, Campus de Luminy, Case 907\\ 13288 Marseille Cedex 09, France}

\date{\small\today}

%%%%%%%%%%%%%%%%%%%%%%%%%%%%%%%%%%%%%
\begin{abstract}\noindent
Group field theory is a generalization of matrix models, with triangulated pseudomanifolds as Feynman diagrams and state sum invariants as Feynman amplitudes. In this paper, we consider Boulatov's three-dimensional model and its Freidel-Louapre positive regularization (hereafter the BFL model) with a `ultraviolet' cutoff, and study rigorously their scaling behavior in the large cutoff limit. We prove an optimal bound on large order Feynman amplitudes, which shows that the BFL model is perturbatively more divergent than the former. We then upgrade this result to the constructive level, using, in a self-contained way, the modern tools of constructive field theory: we construct the Borel sum of the BFL perturbative series via a convergent `cactus' expansion, and establish the `ultraviolet' scaling of its Borel radius. Our method shows how the `sum over triangulations' in quantum gravity can be tamed rigorously, and paves the way for the renormalization program in group field theory.

%Group field theory is a generalization of matrix models, with triangulated pseudomanifolds as Feynman diagrams and state sum invariants as Feynman amplitudes. In this paper, we consider Boulatov's three-dimensional model and its Freidel-Louapre positive regularization (hereafter denoted the BFL model) with a `ultraviolet' cutoff $\Lambda$, and study their perturbative and constructive scaling behavior in the $\Lambda\rightarrow\infty$ limit. Using systematic Cauchy-Schwarz inequalities, we first prove a uniform bound on large order Feynman amplitudes, and describe the class of graphs which saturate it. We then construct the Borel sum of the BFL perturbative series via a convergent `cactus' expansion, and establish the scaling of its Borel radius with the cutoff, confirming the perturbative estimate. Our result shows that the `ultraviolet' behavior of the Boulatov and BFL models differ significantly, and paves the way for the renormalization program in group field theory.
\end{abstract}
%%%%%%%%%%%%%%%%%%%%%%%%%%%%%%%%%%%%%

\maketitle
%\tableofcontents

\section{Introduction}
\subsection{Renormalization in quantum gravity}
%The problem of renormalization in quantum gravity has a long history, made of sweat and tears. 

Since the quantum theory of metric perturbations about Minkowski space-time was proved non-renormalizable in four dimensions \cite{Goroff:1985th}, research in quantum gravity has followed divergent avenues, guided by different reactions to this frustrating no-go result \cite{kiefer}. Supergravity and string theory have relied on extended symmetries and Kaluza-Klein dimensions to tame the ultraviolet (UV) behavior of gravity, even perhaps subsuming it, together with all other particles and interactions, under the unifying concept of strings \cite{Green:1987sp,Polchinski:1998rq}. Non-commutative geometry has explored generalizations of space-time in which UV divergences might disappear, and in which the standard model becomes more natural \cite{Connes:1137119,Connes:2006wp}. The putative existence of a non-Gaussian UV fixed point of quantum general relativity (GR), coined the ``asymptotic safety scenario'' by Weinberg, has been explored in computer simulations, with intriguing results \cite{Niedermaier:2006wt}. Loop quantum gravity (LQG) has followed the historical path, opened by Dirac, of the canonical quantization of the full gravitational field, discarding the issue of renormalizability together with the Minkowski background \cite{Ashtekar:2004eh,Rovelli:2004tv,Thiemann:2007zz}. In other background-independent approaches, such as causal dynamical triangulations \cite{Ambjorn:2006jf} or causets \cite{Henson:2006kf}, the framework of quantum field theory (QFT) is abandoned altogether, and space-time is envisaged as a combinatorial build-up, in the spirit of condensed matter physics.

Even if none of these approaches can claim to have reached maturity, it is already clear that deep and fruitful advances have occurred along the way. Let us mention but a few of them. On the mathematical side, the development of topological quantum field theory has led to major discoveries in the theory of manifolds and knot invariants, initiated by \cite{Witten:1988hf}. On the physical side, the old concept of `space-time discreteness' has become much more precise, to the point where the spectra of area and volume operators are explicitly computable \cite{Rovelli:1994ge,Ashtekar:1996eg,Ashtekar:1997fb}; the non-commutative field theories \cite{Douglas:2001ba} inspired from string theory and noncommutative geometry have `exorcised' the Landau ghost \cite{Disertori:2006nq,Rivasseau:2007ab};  the discovery of the AdS/CFT correspondence has provided a fresh new look on strongly-coupled gauge theory \cite{Aharony:1999ti}; convincing scenarii for the quantum resolution of space-time singularities have been put forward \cite{lrr-2005-11}. And so forth.

%Computer simulations search for RG trajectories of quantum gravity, but numerical work alone cannot explain the fixed points they might see \cite{Comp}. 

In this impressionistic picture of perspectives on quantum gravity, however, a pattern strikes us: the less they rely on background geometric or topological structures, the more remote from standard field-theoretic ideas they are. In particular, the concept of the renormalization group (RG), that is of the hierarchical integration of fluctuations about a computable solution, seems to have somehow faded away.\footnote{Obviously, this is not the case in the asymptotic safety scenario. The recourse to uncontrolled truncations of the effective action, however, makes this program difficult to justify from a mathematical physics perspective.} This Wilsonian paradigm, however, has proved fundamental in many branches of theoretical physics, and is arguably the only encompassing framework to address a system with many interacting degrees of freedom.

Can we find a field-theoretic formulation of quantum gravity, free of {\it a priori} geometric or topological background structure, in which a RG flow could be rigorously defined? In this paper, we take some first steps towards this goal. Our starting point, group field theory (GFT) \cite{Freidel:2005qe,Oriti:2006se}, lies at the crossroads of matrix models (and dynamical triangulations), and LQG, in its spin-foam formulation.

\subsection{Group field theory}

Quantum gravity should implement a sum over all geometries of space-time
with certain weights. From this perspective, general relativity is only of limited guidance: although the Einstein-Hilbert action
suggests how to weigh different metrics on a fixed differential space-time manifold,
it does not tell us how to weigh the sum over different topologies of space-time (or over
differential structures in dimension 4 or above). This problem is loosely called the 
\emph{third quantization} of gravity: quantum gravity may not only be about quantizing metric fluctuations over a classical background topology; perhaps it should also quantize the space-time topology itself.

The discovery that matrix, not scalar models, have a topological content \cite{Hooft:1973jz}
opened a window on how to solve this third quantization problem with field-theoretic methods in two dimensions \cite{David:1985nj}.  In matrix models, Feynman graphs are ribbon graphs, dual to triangulations of surfaces. As the size of the matrix increases, the scaling of observables favors
the particular class of \emph{planar graphs} which correspond to triangulations of the \emph{sphere}, higher-genus surfaces being exponentially suppressed.

The natural candidate generalizations of matrix models in higher dimensions are tensor models, and we propose to nickname $QG_{d}$ the would-be $d$-dimensional model implementing the third quantization of gravity. $QG_{d}$ should follow the rationale of matrix models, namely that space is fruitfully described by gluing together more elementary ``cells" through attaching maps. In the combinatorially simplest case, these cells are just $d$-simplices. Since a $d$-simplex has $d+1$ facets on its boundary, the backbone of our would-be
$QG_d$ should be some abstract $\phi^{d+1}$ interaction on rank-$d$ tensor fields $\phi$. Now, the GFT's of Boulatov and Ooguri \cite{Boulatov:1992vp,Ooguri:1992eb} provide ans\"atze to generate precisely such tensor fields: $\phi$ is a scalar field over $d$ copies of some compact group, and harmonic analysis yields rank-$d$ tensors, which can be naturally coupled in the combinatorial pattern of $d$-simplices. Of course, the exact forms of the propagator and interaction in this setting remain to be found. One should also include a finite list of 
relevant and marginal terms to ensure renormalizability and probably
constructive stability (see below). Ideally, these new terms would also imply (or at least allow
and favor) the standard model matter content of our universe.

Surprisingly enough, LQG, which takes a radically different perspective on quantum gravity, notably by keeping a classical background topology, has converged with the GFT framework. In LQG, the kinematical states of the quantum gravitational field are well understood in terms of spin networks, and the main open issue is the implementation of dynamics, through the Hamiltonian constraint. Spin-foam models provide heuristic definitions of such dynamical transition amplitudes, obtained via a discretization of general relativity on a triangulation of space-time \cite{Baez:1999sr,Perez:2003vx}. There are now interesting such spin-foam models \cite{Freidel:2007py,Engle:2007wy,Livine:2007vk,Livine:2007ya}, hereafter called EPR-FK models,
which in four dimension reproduce Regge gravity in a certain semiclassical limit \cite{Barrett:2009gg,Conrady:2008mk}. There are also some glimpses that they might be just renormalizable \cite{Perini:2008pd}. These spin-foam models, however, capture only a finite subset of the gravitational degrees of freedom, and the question arises of the existence of a `continuum limit'. As the spin-foam amplitude can always be interpreted as a Feynman amplitude of a suitable GFT \cite{Reisenberger:2000fy}, this question boils down to the problem of renormalization in GFT.

This is what makes GFT's exciting: their Feynman amplitudes are spin-foams, which 
implement a tentative background-independent second quantization of gravity,
{\it and} they are scalar field theories of the right tensorial type with the right $\phi^{d+1}$
interaction for third quantization! They thus seem ideally suited for performing both
second and third quantization of gravity in a single bold move. This coincidence also 
leads to an even bolder hope, namely that perhaps adding 
third quantization is just the missing ingredient to the second quantization problem as well.

\subsection{Multiscale renormalization}

%Neither the first versions of string theory, nor spin foams, nor non-commutative geometry, nor computer studies of gravity really attack head-on third quantization. Only
%group field theory and perhaps M-theory does \cite{Fgft,Ogft1,Ogft2}.

%most notably supergravity, string theory, non-commutative geometry, loop quantum gravity and its most advanced version, namely group field theory, and numerical simulations.

To begin the program of renormalization of GFT, the first step is the definition of scales. However, because third quantization blurs space-time, it fools our deeply ingrained relation of scale
to distances (centimeter scale, millimeter scale...). In particular our familiar intuition that ultraviolet is about short distance and infrared about long distance effects may no longer hold. 
Fortunately non-commutative field theory, where ultraviolet-infrared mixing can occur, as in the Grosse-Wulkenhaar model \cite{Grosse:2004yu}, has familiariazed us with such conceptual issues. It provides a hint as to the extended definition of scales and renormalization group (RG) which are presumably required for $QG_d$ \cite{Rivasseau:2003zy}.

Roughly speaking, to set up a RG analysis in QFT, one must draw a line between propagator and interaction. There should be no scales in the interaction, because vertices 
should always be considered, by definition, the correct ``local" objects (even when, like in the $x$-space representation of the GW model, they naively don't look local at all). 
Hence scales should be entirely contained in the propagator,
more precisely in its spectrum, which does not depend on the basis used. 
In standard Euclidean QFT, a propagator is a positive covariance $C$ which is the 
inverse of a positive Hermitian operator $H$, and we are led to define a physical scale as 
a range of eigenvalues for $H$. A RG trajectory 
should then integrate over `ultraviolet' scales (large eigenvalues) towards `infrared' scales (smaller eigenvalues). The best way to cut the typical geometrically growing sequence of RG slices
is through the parametric representation of the propagator as
\be  C= H^{-1} = \int_0^{\infty} e^{- \alpha H} d \alpha = \sum_{i=0}^{\infty} 
C_i, C_0 = \int_{1}^{\infty}e^{- \alpha H} d \alpha ,\;\; 
C_i = \int_{M^{-2i}}^{M^{-2(i-1)}}e^{- \alpha H} d \alpha \;\;  {\rm for} \;\; i\ge 1,
\ee 
where $M$ is a fixed number greater than 1\footnote{The metric system uses $M=10$ for 
reasons linked to human anatomy, but computers might prefer $M=2$
and mathematicians $M=e$.}. $C_i$ should be considered the definition of a (discrete) RG scale.
The RG trajectory then performs functional integration over scales, one at a time.
A renormalizable trajectory is one in which the relevant part of the RG flow changes
only a finite number of terms. In beautiful cases like QCD, such a trajectory joins
a perturbatively computable UV fixed point (asymptotic freedom) to a non-trivial 
phase transition (quark confinement). We hope for such a scenario, namely that a consistent $QG_4$ flow could join some UV fixed point (presumably topological or at least with enhanced symmetry) to a non-trivial phase transition at which our ordinary effective geometry would emerge.\footnote{Such a transition is usually called \emph{geometrogenesis}. But 
even a $QG_4$ ultraviolet topological fixed point might not be the end of physics. It could hide
at still higher (transplanckian?) phase transition
which we could e.g. nickname \emph{topologicogenesis},
perhaps as a condensation of a combinatoric phase. 
Quantum graphity is a candidate catchy name for such a combinatoric phase \cite{Konopka:2006hu}. 
But such a highly speculative scenario presumably cannot be explored with $QG_4$ alone.
In our notation, graphity or such meta-theories of gravity could be nicknamed $QG_{\infty}$.}

Within this program, we hope to clarify the meaning of renormalizability for third quantized theories, and in particular to answer whether or not the recent EPR-FK models define a renormalizable quantum theory of gravity, to understand how they should be physically interpreted, and eventually to find out whether and how they should be modified for full 
mathematical consistency up to the constructive level.  

This is an ambitious program and it is reasonable to gain better expertise first on 
$QG_3$ before tackling $QG_4$. The leading candidate for a
$QG_3$ theory is the Boulatov model with gauge group $\su$. Its 
amplitudes are those of the Ponzano-Regge model \cite{PR} and like matrix models
it is topological. The recent identification of a class of graphs which
for 3d tensorial models generalizes 2d planar graphs, in the sense of having the highest superficial degree of divergence  \cite{Freidel:2009hd}, is an important 
step for this program. With this paper, we address the question of power counting from a slightly different perspective, focusing on general bounds rather than explicit estimates for particular classes of graphs. 

%In \cite{FGO} a power counting is established for a particular class of graphs called type I
%graphs which is expected to dominate as the cutoff $\Lambda$ gets large.
%They are therefore the 3D generalization of planar (one particle irreducible) graphs in 2D. However the amplitudes of non-type I graphs are neither evaluated nor bounded in \cite{FGO}, so that the dominance of type I graph is not proved. It is one purpose of this paper to provide a uniform perturbative  bound on the graphs of the Boulatov model which is in a way optimal since it is  saturated for a particular class of type-I graphs. This brings strong support to the conjectures of \cite{FGO}.  

\subsection{Constructive field theory}

Constructive field theory was initially the name of a mathematical physics program launched forty years ago by Arthur Wightman to define rigorously  particular field theory models  in increasing order of difficulty, and check that they obeyed the ``Wightman axioms" \cite{Streater:1989vi}. 

The philosophy was to introduce as many cutoffs as necessary for the quantum field 
correlation functions of these models to be well defined, and then to develop 
the necessary methods to lift these cutoffs. But ordinary perturbation theory could not be used
directly since for Bosonic field theories  it \emph{diverges}, i.e. it has zero radius of convergence.  In constructive theory,
the divergence of perturbation series is addressed with the same care 
as the UV divergence of individual Feynman amplitudes.
Techniques had to be developed to rewrite the formal expansions of ordinary quantum field theory as {\it convergent} series. Certainly the renormalization group of Wilson is the most 
powerful such tool. It had to be adapted to constructive purposes \cite{Riv} and was rewritten 
in the form of rigorous multiscale expansions,
slowly improved over the years in many ways, and generalized to new situations, 
such as the RG around the Fermi surface of condensed matter \cite{Rivasseau:1995mm}.

Early constructive field theory succeeded in building rigorously super-renormalizable
field theories such as the emblematic $\phi^4_2$ or $\phi^4_3$ models. It also elucidated
their relationship to perturbation theory: the Schwinger functions of 
these models are the Borel sum of their perturbation theory \cite{EMS,MS}. But $\phi^4_4$ itself could not be built, since its coupling constant does not remain small in the ultraviolet regime.
Being asymptotically free, non-Abelian gauge theories do not have this problem.
Nevertheless, although some partial results were obtained, they could not be built in the full constructive sense either, due to technical difficulties such as Gribov ambiguities; neither could the interesting infrared confining regime of the theory be understood rigorously.
Probably the first 4 dimensional field theory that will be built 
completely through constructive methods will be the Grosse-Wulkenhaar model \cite{Grosse:2004yu},
a non-commutative field theory which, ironically, should not satisfy
the Wightman axioms of the initial constructive program.

The constructive program is largely unknown in the quantum gravity community, with the notable exception
of \cite{Freidel:2002tg}. However, we
think that constructive theory embodies a deeper point of view on QFT than the usual one.
A modern constructive technique such as the cactus expansion used in this paper allows to resum perturbation theory by reorganizing it in a precise, explicit
though completely different manner than the usual Feynman graphs. Rather than resumming families of entire Feynman amplitudes, this constructive recipe splits Feynman amplitudes
into lots of finer pieces, according to a precise rule involving new interpolating parameters, and resum
infinite families of these finer pieces.
In the end the cactus expansion itself is a sum of new ``amplitudes" indexed by trees or ``cacti".
%\footnote{These ``constructive trees" or cacti are similar to the trees of a cluster expansion and should not be confused  with the ``Gallavotti-Nicol\`o"
%or ``Connes-Kreimer``'' trees, which are related to Zimmermann's forests. Nodes of
%the latter correspond to subgraphs of Feynman graphs.
%However there is a link: the essence of constructive multiscale analysis is to require compatibility between these two 
%different sorts of trees; the constructive trees must
%be chosen so that their restriction to each node of the ``Gallavotti trees" must be a 
%subtree spanning that node.} 
The series obtained in this way \emph{converges absolutely}. We believe that
this convergent expansion (which is no longer a power series in the coupling constant) should 
be considered the mathematically correct version of perturbation theory.\footnote{From this perspective, 
one should not confuse the words ``constructive" and ``non-perturbative". In field theory
the word non-perturbative usually refers to effects which cannot be seen in perturbation theory,
such as instanton effects, or to the simulation of functional integrals e.g. through Monte-Carlo computations. In contrast, constructive theory is a convergent reorganization of perturbation theory  and is not yet a theory of non-perturbative effects, which, to this day, remains
to be mathematically developed.}

Introducing these constructive methods for more exotic theories such as the GFT's
should bring further constraints on the mathematical consistency of these theories. 
These additional constraints are precious in view of our lack of experiments to test the various
theories of quantum gravity.

\subsection{Plan of the paper}

The plan of our paper is as follows. Section II introduces the Boulatov 
and Boulatov-Freidel-Louapre (BFL) models. In section III we establish a set of uniform perturbative bounds, and exhibit classes of graphs which saturate them, showing that the Boulatov and BFL models
have very different `ultraviolet' behavior.  In section IV we introduce the constructive `cactus' expansion, 
which allows to construct the Borel sum of a general $\phi^4$ model. In section V, we perform this construction for the BFL model, and obtain a bound for the Taylor-Borel remainders of the theory which confirms the perturbative estimate. We present our 
conclusions and future program in section VI.

Our main results are Theorems \ref{borelBFL}, \ref{pertB}, \ref{pertBFL} and \ref{satur}. 
They establish rigorously that the Boulatov model (\emph{without mass renormalization}) 
scales perturbatively as $\Lambda^{3/2}$ per vertex at large cutoff $\Lambda$ and that
the BFL model scales both perturbatively and constructively as $\Lambda^{3}$ 
per vertex at large cutoff $\Lambda$ -- which is much worse. 

The main technique for both perturbative and constructive bounds used in this
paper is the Cauchy-Schwarz inequality, applied to appropriate cuts of the lines and/or the vertices
of the graph. This is also the main tool used in \cite{Abdesselam} in a study of the volume conjecture for
classical spin networks, a program which has some obvious overlap with ours.
%%%%%%%%%%%%%%
\section{The regularized Boulatov model}
\subsection{Boulatov's original model}
\begin{figure}[h]
\scalebox{0.8} % Change this value to rescale the drawing.
{
\begin{pspicture}(0,-2.4)(15.696783,2.4)
\psbezier[linewidth=0.024](7.496783,1.92)(7.496783,1.12)(7.496783,0.72)(6.2967825,0.72)
\psbezier[linewidth=0.024](8.2967825,-1.28)(8.2967825,-0.48)(8.2967825,-0.08)(9.496782,-0.08)
\psbezier[linewidth=0.024](9.496782,0.72)(8.696783,0.72)(8.2967825,0.72)(8.2967825,1.92)
\psbezier[linewidth=0.024](6.2967825,-0.08)(7.0967827,-0.08)(7.496783,-0.08)(7.496783,-1.28)
\psline[linewidth=0.024cm](7.8967824,1.92)(7.8967824,-1.28)
\psline[linewidth=0.024cm](6.2967825,0.32)(7.6967826,0.32)
\psline[linewidth=0.024cm](8.096783,0.32)(9.496782,0.32)
\psbezier[linewidth=0.024](13.096783,1.92)(13.096783,1.12)(13.096783,0.72)(11.896783,0.72)
\psbezier[linewidth=0.024](15.096783,0.72)(14.2967825,0.72)(13.896783,0.72)(13.896783,1.92)
\psbezier[linewidth=0.024](11.896783,-0.08)(12.696783,-0.08)(13.096783,-0.08)(13.096783,-1.28)
\psbezier[linewidth=0.024](13.896783,-1.28)(13.896783,-0.48)(13.896783,-0.08)(15.096783,-0.08)
\psbezier[linewidth=0.024](11.896783,0.32)(13.696783,0.32)(13.496782,1.12)(13.496782,1.92)
\psbezier[linewidth=0.024](13.496782,-1.28)(13.496782,0.52)(14.096783,0.32)(15.096783,0.32)
\usefont{T1}{ptm}{m}{n}
\rput(7.976783,-2.235){Tetrahedron}
\usefont{T1}{ptm}{m}{n}
\rput(13.5467825,-2.215){Pillow}
\usefont{T1}{ptm}{m}{n}
\rput(7.5167828,2.16){\footnotesize $1$}
\usefont{T1}{ptm}{m}{n}
\rput(7.9167824,2.16){\footnotesize $2$}
\usefont{T1}{ptm}{m}{n}
\rput(8.316783,2.16){\footnotesize $3$}
\usefont{T1}{ptm}{m}{n}
\rput(9.736783,0.72){\footnotesize $3$}
\usefont{T1}{ptm}{m}{n}
\rput(9.736783,0.32){\footnotesize $4$}
\usefont{T1}{ptm}{m}{n}
\rput(9.736783,-0.04){\footnotesize $5$}
\usefont{T1}{ptm}{m}{n}
\rput(8.2967825,-1.5){\footnotesize $5$}
\usefont{T1}{ptm}{m}{n}
\rput(7.8767824,-1.5){\footnotesize $2$}
\usefont{T1}{ptm}{m}{n}
\rput(7.5167828,-1.5){\footnotesize $6$}
\usefont{T1}{ptm}{m}{n}
\rput(6.0767827,0.7){\footnotesize $1$}
\usefont{T1}{ptm}{m}{n}
\rput(6.0767827,0.3){\footnotesize $4$}
\usefont{T1}{ptm}{m}{n}
\rput(6.0767827,-0.06){\footnotesize $6$}
\usefont{T1}{ptm}{m}{n}
\rput(13.116782,2.22){\footnotesize $1$}
\usefont{T1}{ptm}{m}{n}
\rput(13.516783,2.22){\footnotesize $2$}
\usefont{T1}{ptm}{m}{n}
\rput(13.916782,2.22){\footnotesize $3$}
\usefont{T1}{ptm}{m}{n}
\rput(15.416782,0.7){\footnotesize $3$}
\usefont{T1}{ptm}{m}{n}
\rput(15.416782,0.3){\footnotesize $4$}
\usefont{T1}{ptm}{m}{n}
\rput(15.416782,-0.06){\footnotesize $5$}
\usefont{T1}{ptm}{m}{n}
\rput(11.656782,0.68){\footnotesize $1$}
\usefont{T1}{ptm}{m}{n}
\rput(11.656782,0.28){\footnotesize $2$}
\usefont{T1}{ptm}{m}{n}
\rput(11.656782,-0.08){\footnotesize $6$}
\usefont{T1}{ptm}{m}{n}
\rput(13.936783,-1.5){\footnotesize $5$}
\usefont{T1}{ptm}{m}{n}
\rput(13.516783,-1.5){\footnotesize $4$}
\usefont{T1}{ptm}{m}{n}
\rput(13.156782,-1.5){\footnotesize $6$}
\psline[linewidth=0.024cm](0.012,0.8)(1.5967826,0.8)
\psline[linewidth=0.024cm](0.012,0.4)(1.5967826,0.4)
\psline[linewidth=0.024cm](0.012,0.0)(1.5967826,0.0)
\psframe[linewidth=0.024,dimen=outer](2.1967826,1.0)(1.5967826,-0.2)
\psline[linewidth=0.024cm](2.1967826,0.8)(3.7967825,0.8)
\psline[linewidth=0.024cm](2.1967826,0.4)(3.7967825,0.4)
\psline[linewidth=0.024cm](2.1967826,0.0)(3.7967825,0.0)
\usefont{T1}{ptm}{m}{n}
\rput(1.9167826,-2.235){Covariance}
\end{pspicture} 
}
\caption{The covariance $C_{\phi}$, the Boulatov tetrahedral vertex $T[\phi]$, and the Freidel-Louapre pillow $P[\phi]$. The labels on the vertices match the ordering of the group elements in the integrand of $T[\phi]$ and $P[\phi]$.}
\label{tetrahedron}
\end{figure}
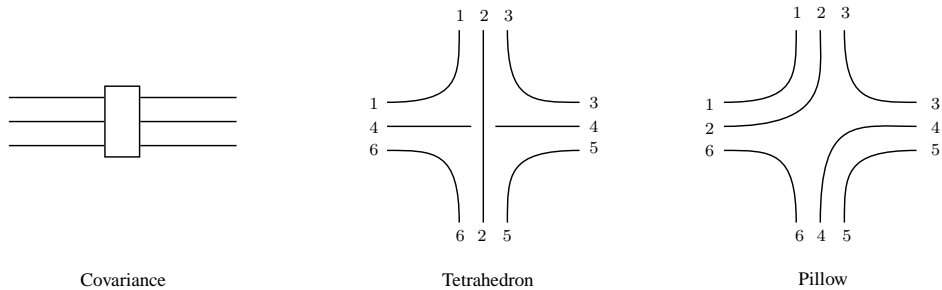

Boulatov's model \cite{Boulatov:1992vp} is a group field theory whose dynamical variable is a real-valued function $\phi$ over the group $\su^3$.
Furthermore, the field is required to be invariant under the right diagonal action of $\su$ and cyclic permutations $c$ of its arguments, i.e.:
\be\label{invariances}
\phi(g_1h,g_2h,g_3h)=\phi(g_1,g_2,g_3) \;\;\;\;\;\;, \;\;\;\;\phi(g_{c(1)},g_{c(2)},g_{c(3)})=\phi(g_1,g_2,g_3).
\ee
The ``dynamics" of the field is governed by the following non-local action:
\be
S_B[\phi]:=\f{1}{2}\int\prod_{i=1}^3dg_i\ \phi^2(g_1,g_2,g_3)+\f{\lambda}{8}\int\prod_{i=1}^{6}dg_i\ \phi(g_1,g_2,g_3)\phi(g_3,g_4,g_5)\phi(g_5,g_2,g_6)\phi(g_6,g_4,g_1),
\ee
where $dg_i$ denotes the $\su$ Haar measure. 
Note that the quadratic term is a kind of pure mass term with no ``kinetic" component.
In the quartic term, denoted $T[\phi]$ below, the six integration variables are repeated twice, following the pattern of the edges of a tetrahedron, as illustrated in Fig. \ref{tetrahedron}. (The factor $8$ is only for simpler final formulas.)

In the language of degenerate Gaussian measures (Appendix A), 
the Boulatov partition function can be formally defined as 
\be\label{formal}
\mathcal{Z}_B(\lambda):=\int d\mu[\phi]\ e^{-\lambda T[\phi]/8},
\ee
where $d\mu[\phi]$ is the normalized Gaussian measure whose covariance $C_{\phi}$ is given by the symmetrizer 
\be
(C_{\phi}\phi)(g_1,g_2,g_3):=\f{1}{3}\sum_c\int dh\ \phi(g_{c(1)}h,g_{c(2)}h,g_{c(3)}h).
\ee
Thus, in (\ref{formal}), the field $\phi$ is generic, and the invariances (\ref{invariances}) are implemented by the orthogonal projector $C_{\phi}$.

%The sum runs over the 3 cyclic permutations $c$ and the factor $1/3$ appears for trivial normalization purposes. 

\subsection{A positive regularization}

The partition function (\ref{formal}) is only formal and needs regularizations to become mathematically well-defined. The problem is twofold: the Fourier space of the field $\phi$ is non-compact, although discrete, and hence `ultraviolet' divergences arise; Boulatov's quartic interaction $T[\phi]$ is not positive, or unstable, and is thus unsuited for constructive considerations. 

To cure the first problem, we follow \cite{Freidel:2002tg} and introduce an ultra-violet cutoff $\La$ truncating the Peter-Weyl (or Fourier) decomposition of the field:
\be
\phi(g_1,g_2,g_3)=\sum_{{j}_1,j_2,j_3}^{\La} \text{tr} \left(\Phi_{j_1,j_2,j_3} D^{j_1}(g_1) D^{j_2}(g_2) D^{j_3}(g_3)\right).
\ee
In this formula, the sum runs over the spins $j_1,j_2,j_3$ up to $\La$; $D^j(g)$ denotes the $(2j+1)$-dimensional matrix representation of $g$; $\Phi_{j_1,j_2,j_3}$ are the Fourier modes of the field $\phi$
viewed as complex-valued tensors and $\text{tr}$ denotes the trace in the space carrying the tensor product representation associated to the spins $j_{1}$, $j_{2}$, $j_{3}$. In the following, we denote $\cH^{(\La)}$ the subspace of $L^2(\su^3)$ resulting from this truncation, and $\cH_{0}^{(\La)}:=\cH^{(\La)}\cap\text{Im}\ C_{\phi}$. The number of degrees of freedom left is thus given by

\be
\dim\cH_{0}^{(\La)}=\mathcal{O}(\La^6).
\ee

Now the second problem. That Boulatov's interaction $T[\phi]$ is not positive can be seen from its Fourier space formulation where the interaction term
reduces to an oscillatory $\{6j\}$ symbol \cite{Boulatov:1992vp}. 
To fix this shortcoming, Freidel and Louapre propose to add the following `pillow' term\footnote{The word `pillow' refers to the geometric interpretation of the GFT vertex: if Boulatov's $T[\phi]$ is a tetrahedron, then Freidel and Louapre's $P[\phi]$ are two tetrahedra glued along two triangles -- a pillow.} to the action \cite
{Freidel:2002tg}
\be
P[\phi]:=\int\prod_{i=1}^{6}dg_i\ \phi(g_1,g_2,g_3)\phi(g_3,g_4,g_5)\phi(g_5,g_4,g_6)\phi(g_6,g_2,g_1).
\ee
Indeed, they show that when $\vert\delta\vert\leq1$,  $I_{\delta}[\phi]:=P[\phi]+\delta T[\phi]$ is positive. To this aim, they introduce the `squaring' operator $S$ mapping $\phi$ to the function $S\phi$ on $\su^4$ defined by
 $S\phi(g_1,g_2,g_3,g_4):=\int dg\ \phi(g_1,g_2,g)\phi(g,g_3,g_4)$.
In terms of this new field, the modified interaction $I_{\delta}$ reads
\be
I_{\delta}[\phi]=\la S\phi\vert(1+\delta\mathcal{T}) S\phi\ra_4,
\ee
where $\mathcal{T}$ is the involution transposing the central arguments of $S\phi$:
$\mathcal{T}S\phi(g_1,g_2,g_3,g_4):=S\phi(g_1,g_3,g_2,g_4)$, and $\la\cdot\vert\cdot\ra_{4}$ is the standard inner product in $L^2(\su^4)$. $(1+\delta\mathcal{T})$ being a positive operator, the modified quartic interaction $I_{\delta}$ is clearly positive. 

% Generated with LaTeXDraw 2.0.2
% Mon Jun 29 13:00:25 CEST 2009
% \usepackage[usenames,dvipsnames]{pstricks}
% \usepackage{epsfig}
% \usepackage{pst-grad} % For gradients
% \usepackage{pst-plot} % For axes

Combining the cutoff on spins $\La$ and the Boulatov-Freidel-Louapre (BFL) interaction, we get the, now well-defined, regularized partition function
\be
\mathcal{Z}^{(\La)}_{\text{BFL}}(\lambda):=\int d\mu^{(\La)}[\phi]\ e^{-\lambda I_{\delta}[\phi]/8}.
\ee

In \cite{Freidel:2002tg}, Freidel and Louapre showed that $\mathcal{Z}^{(\La)}_{\text{BFL}}$ is Borel summable. Their method, however, does not allow to prove Borel summability of the (more physically relevant) free energy $\cF^{(\La)}_{\text{BFL}}:=\log \mathcal{Z}^{(\La)}_{\text{BFL}}$, and provides no control over the $\La\rightarrow\infty$ limit. In this paper, we construct $\cF^{(\La)}_{\text{BFL}}$ explicitly, and determine the scaling behaviour of its Borel radius as $\La\rightarrow\infty$. But before facing these constructive considerations, let us pause a moment to establish the perturbative power counting of the Boulatov and BFL models.

\section{Perturbative Bounds}

Let $\lambda^n\cA_G$ denote the amplitude of an order-$n$ Feynman graph $G$ for these models. In this section we prove the following

\begin{theorem}\label{pertB}
There exists a constant  $K$ such that for any connected Boulatov vacuum graph $G$ of order $n$
\be  \vert\cA^B_G\vert  \leq K^n \Lambda^{6 + 3n/2}.
\ee
\end{theorem}

\begin{theorem}\label{pertBFL}
There exists a constant  $K$ such that for any connected BFL vacuum graph $G$ of order $n$
\be  \vert\cA^{BFL}_G \vert \leq K^n \Lambda^{6 + 3n}.
\ee
\end{theorem}

\begin{theorem}\label{satur}
These bounds are optimal in the sense that 
there exist graphs $G_n$ and $G'_{n}$ of order $n$ 
such that $ \cA^{B}_{G'_n}  \simeq K^n \Lambda^{6 + 3n/2}$ and $ \cA^{BFL}_{G_n}  \simeq K^n \Lambda^{6 + 3n}$.
\end{theorem}

Here, we focus only on vacuum graphs, but these results can easily be extended to the general case.

Let $G$ be an amputated $\phi^4$ graph, i.e. a graph with coordination 4 at each inner vertex
and coordination less than 4 at its outer vertices. We say that a set $A$ of vertices 
of $G$ is connected if the subgraph made of these vertices and all their inner lines (that is all lines of $G$ starting 
and ending at a vertex of $A$) is connected. We collect
now some graph theoretic definitions that will be useful below.

\begin{definition}
A generalized tadpole is a two-point graph with only one external vertex.
\end{definition}
\begin{definition}
An $(A,B)$-cut of a two-point connected graph $G$ with two external vertices $v_A$ and $v_B$
is a partition of the vertices of $G$ into two subsets $A$ and $B$
such that $v_A \in A$, $v_B \in B$ and $A$ and $B$ are connected.
\end{definition}
\begin{definition}
A line joining a vertex of $A$ to a vertex of $B$ in the graph is called a frontier line
for the $(A,B)$-cut. 
A vertex of $B$ is called a frontier vertex with respect to the cut if there is a frontier
line attached to that vertex.
\end{definition}
Remark by parity that there must be an odd number of such frontier lines.
\begin{definition}
An exhausting sequence of cuts for a connected  two-point graph $G$ of order $n$
is a sequence $A_0 =\emptyset  \varsubsetneqq
A_1  \varsubsetneqq A_2 \varsubsetneqq \cdots  \varsubsetneqq A_{n-1}  \varsubsetneqq A_n =G$
such that $(A_p, B_p := G\setminus A_p)$ is a cut of $G$ for any $p=1,  \cdots , n-1$. 
\end{definition}
Hence an exhausting sequence of cuts is a total ordering of the vertices of $G$, such that each vertex 
is roughly speaking pulled successively through the frontier without disconnecting $A$ nor $B$.

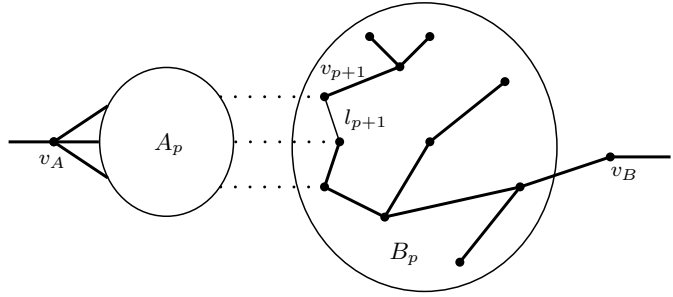
\begin{figure}[h]
\begin{center}
\scalebox{1} % Change this value to rescale the drawing.
{
\begin{pspicture}(0,-1.93)(8.82,1.93)
\psdots[dotsize=0.12](0.6,0.07)
\psdots[dotsize=0.12](0.6,0.07)
\psline[linewidth=0.04cm](0.0,0.07)(0.6,0.07)
\psline[linewidth=0.04cm](0.6,0.07)(1.32,0.55)
\psline[linewidth=0.04cm](0.6,0.07)(1.2,0.07)
\psline[linewidth=0.04cm](0.6,0.07)(1.32,-0.41)
\psellipse[linewidth=0.02,dimen=outer](2.1,0.07)(0.9,1.0)
\usefont{T1}{ptm}{m}{n}
\rput(2.13,0.035){$A_p$}
\psellipse[linewidth=0.02,dimen=outer](5.53,0.0)(1.77,1.93)
\usefont{T1}{ptm}{m}{n}
\rput(0.58,-0.19){\footnotesize $v_A$}
\usefont{T1}{ptm}{m}{n}
\rput(8.19,-0.35){\footnotesize $v_B$}
\psdots[dotsize=0.12](8.0,-0.13)
\psline[linewidth=0.02cm](4.2,0.67)(4.4,0.07)
\psline[linewidth=0.04cm](4.4,0.07)(4.2,-0.53)
\psline[linewidth=0.04cm](4.2,-0.53)(5.0,-0.93)
\psline[linewidth=0.04cm](5.0,-0.93)(6.8,-0.53)
\psline[linewidth=0.04cm](5.0,-0.93)(5.6,0.07)
\psline[linewidth=0.04cm](5.6,0.07)(6.6,0.87)
\psline[linewidth=0.04cm](6.8,-0.53)(8.0,-0.13)
\psline[linewidth=0.04cm](6.8,-0.53)(6.0,-1.53)
\psdots[dotsize=0.12](6.8,-0.53)
\psdots[dotsize=0.12](6.0,-1.53)
\psdots[dotsize=0.12](5.0,-0.93)
\psdots[dotsize=0.12](5.6,0.07)
\psdots[dotsize=0.12](6.6,0.87)
\psdots[dotsize=0.12](4.2,-0.53)
\psdots[dotsize=0.12](4.4,0.07)
\psdots[dotsize=0.12](4.2,0.67)
\psline[linewidth=0.04cm](4.2,0.67)(5.2,1.07)
\psline[linewidth=0.04cm](5.2,1.07)(4.8,1.47)
\psline[linewidth=0.04cm](5.2,1.07)(5.6,1.47)
\psdots[dotsize=0.12](5.2,1.07)
\psdots[dotsize=0.12](4.8,1.47)
\psdots[dotsize=0.12](5.6,1.47)
\psline[linewidth=0.04cm,linestyle=dotted,dotsep=0.16cm](2.8,0.67)(4.2,0.67)
\psline[linewidth=0.04cm,linestyle=dotted,dotsep=0.16cm](3.0,0.07)(4.4,0.07)
\psline[linewidth=0.04cm,linestyle=dotted,dotsep=0.16cm](2.8,-0.53)(4.2,-0.53)
\usefont{T1}{ptm}{m}{n}
\rput(4.45,0.99){\footnotesize $v_{p+1}$}
\usefont{T1}{ptm}{m}{n}
\rput(4.75,0.37){\footnotesize $l_{p+1}$}
\psline[linewidth=0.04cm](8.0,-0.13)(8.8,-0.13)
\usefont{T1}{ptm}{m}{n}
\rput(5.26,-1.425){$B_p$}
\end{pspicture} 
}
\end{center}
\caption{$(A_{p},B_{p})$-cut in a two-point graph: the dotted lines are the frontier lines, and the solid lines in $B_{p}$ represent the spanning tree $T_{p}$, the remaining lines (loop lines) being omitted. When the thinner line $l_{p+1}$ is deleted, $T_{p}$ splits into $R_{p}$ (bottom) and $S_{p}$ (top).}
\label{trees}
\end{figure}

\begin{lemma}
If $G$ is a connected  two-point graph which has no generalized tadpoles subgraphs $S \subset G$, 
there exists an exhausting sequence of cuts for $G$.
\end{lemma}
\prf $\ $  We proceed by induction. We suppose such a sequence $A_0 =\emptyset  \varsubsetneqq
A_1  \varsubsetneqq A_2 \varsubsetneqq \cdots A_p$ has been built for $0\le p<n-1$. 
We want to find a frontier vertex $v_{p+1}$, such that $A_{p+1}=A_{p}\cup\{v_{p+1}\}$. We consider a rooted tree $T_p$
spanning $B_p$, hence with $(n-p)-1$ lines, with root $v_B$ (see Fig. \ref{trees}).

This rooted tree induces a partial ordering of the vertices of $B_p$.
Since $B_p$ is finite  there exists a maximal frontier
vertex $v_{max} $ with respect to that ordering,
that is a frontier vertex such that the  ``branch above $v_{max}$" in $T_p$ does not contain any 
other frontier vertex.

We notice first that $v_{max} \ne v_B$. Indeed, otherwise it would mean $v_B$ is the only frontier vertex
left in $B_p$. The number of frontier lines being odd, either $v_B$ would have three frontier lines which would mean
$B_p = \{v_B\}$ and would contradict $p < n-1$, or it would have one frontier line and there would
be a generalized tadpole at $v_B$.

Second, we claim that picking $v_{p+1} = v_{max}$ is a valid choice, namely
that $(A_p \cup\{v_{p+1}\}, B_p\setminus \{v_{p+1}\}  )$ is a cut. It is clear that $A_p \cup\{v_{p+1}\}$
is connected. So it remains to check that $B_{p+1} = B_p \setminus\{v_{p+1}\}$ is still connected through its inner lines. 
Call $\ell_{p+1}$ the unique line hooked to $v_{p+1}$ in the path in $T_p$ from $v_{p+1}$ to the root $v_B$.
Cutting $\ell_{p+1}$ splits $T_p$ into two connected components, one of which, $R_p$, contains $v_B$,
and the other is a rooted tree $S_p$ with root and only frontier vertex $v_{p+1}$. Since $v_{p+1}$ is a frontier vertex, it has at most three lines in $B_p$, hence
there are at most two lines from $v_{p+1}$ to $B_{p+1}$ distinct from $\ell_{p+1}$. 
The tree $R_p$ does not contain any line hooked to $v_{p+1}$, hence its lines remain inner
lines of $B_{p+1}$ so all the vertices in $R_p$ remain in a single connected component of $B_{p+1}$.
If $B_{p+1}$ is not connected, this would mean, first, that $S_p$ must contain other vertices
than $v_{p+1}$ and, second, that 
removing in $S_p$ its root $v_{p+1}$ and the (at most two) lines of $S_p$ hooked to it,
we would obtain one or two connected components, made of the vertices of $S_p\setminus \{v_{p+1}\}$
plus their inner lines, which no longer hook to $R_p$ through inner lines of $B_{p+1}$. 
Since these components have no frontier vertices hence no frontier lines, and since there are 
no 1-point subgraphs in the $\phi^4$ theory,
it would mean that this component is in fact unique and must have been hooked to $G$ through 
exactly two lines  from $v_{p+1}$ to $B_{p+1}$ distinct from $\ell_{p+1}$, hence it would 
have been a generalized tadpole.
\qed

% Generated with LaTeXDraw 2.0.2
% Fri Jun 26 15:28:01 CEST 2009
% \usepackage[usenames,dvipsnames]{pstricks}
% \usepackage{epsfig}
% \usepackage{pst-grad} % For gradients
% \usepackage{pst-plot} % For axes

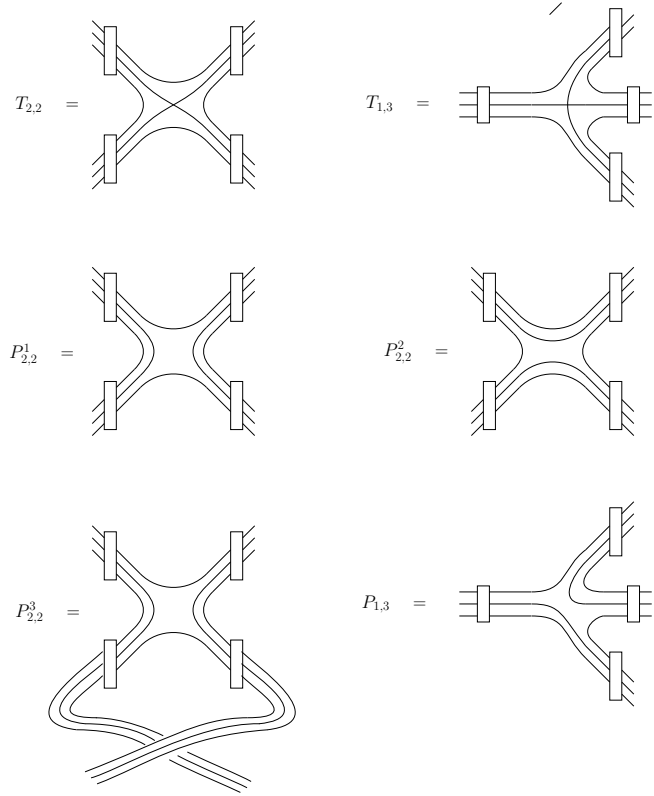
\begin{figure}[h]
\begin{center}
\scalebox{0.4} % Change this value to rescale the drawing.
{
\begin{pspicture}(0,-13.149745)(22.649712,13.149745)
\psline[linewidth=0.03cm](4.84,11.734744)(5.64,10.934745)
\psline[linewidth=0.03cm](4.84,11.334744)(5.64,10.534744)
\psline[linewidth=0.03cm](4.84,10.934745)(5.44,10.334744)
\psframe[linewidth=0.03,dimen=middle](4.84,12.334744)(4.44,10.734744)
\psline[linewidth=0.03cm](4.04,12.534744)(4.44,12.134745)
\psline[linewidth=0.03cm](4.04,12.134745)(4.44,11.734744)
\psline[linewidth=0.03cm](4.44,11.334744)(4.04,11.734744)
\psline[linewidth=0.03cm](8.64,11.734744)(7.84,10.934745)
\psline[linewidth=0.03cm](8.64,11.334744)(7.84,10.534744)
\psline[linewidth=0.03cm](8.64,10.934745)(8.04,10.334744)
\psframe[linewidth=0.03,dimen=middle](9.04,12.334744)(8.64,10.734744)
\psline[linewidth=0.03cm](9.44,12.534744)(9.04,12.134745)
\psline[linewidth=0.03cm](9.44,12.134745)(9.04,11.734744)
\psline[linewidth=0.03cm](9.04,11.334744)(9.44,11.734744)
\psline[linewidth=0.03cm](4.84,7.7347445)(5.64,8.534744)
\psline[linewidth=0.03cm](4.84,8.134745)(5.64,8.934745)
\psline[linewidth=0.03cm](4.84,8.534744)(5.44,9.134745)
\psframe[linewidth=0.03,dimen=middle](4.84,8.734744)(4.44,7.1347446)
\psline[linewidth=0.03cm](4.04,6.9347444)(4.44,7.3347445)
\psline[linewidth=0.03cm](4.04,7.3347445)(4.44,7.7347445)
\psline[linewidth=0.03cm](4.44,8.134745)(4.04,7.7347445)
\psline[linewidth=0.03cm](8.64,7.7347445)(7.84,8.534744)
\psline[linewidth=0.03cm](8.64,8.134745)(7.84,8.934745)
\psline[linewidth=0.03cm](8.64,8.534744)(8.04,9.134745)
\psframe[linewidth=0.03,dimen=middle](9.04,8.734744)(8.64,7.1347446)
\psline[linewidth=0.03cm](9.44,6.9347444)(9.04,7.3347445)
\psline[linewidth=0.03cm](9.44,7.3347445)(9.04,7.7347445)
\psline[linewidth=0.03cm](9.04,8.134745)(9.44,7.7347445)
\psbezier[linewidth=0.03](5.44,10.334744)(5.84,9.934745)(5.84,9.534744)(5.44,9.134745)
\psbezier[linewidth=0.03](8.04,10.334744)(7.64,9.934745)(7.64,9.534744)(8.04,9.134745)
\psbezier[linewidth=0.03](5.64,10.934745)(6.24,10.334744)(7.24,10.334744)(7.84,10.934745)
\psbezier[linewidth=0.03](5.64,8.534744)(6.24,9.134745)(7.24,9.134745)(7.84,8.534744)
\psline[linewidth=0.03cm](17.24,10.134745)(18.628958,10.134745)
\psline[linewidth=0.03cm](17.24,9.734744)(18.628958,9.734744)
\psline[linewidth=0.03cm](17.24,9.334744)(18.628958,9.334744)
\psline[linewidth=0.03cm](21.24,12.334744)(20.44,11.534744)
\psline[linewidth=0.03cm](21.24,11.934745)(20.44,11.134745)
\psline[linewidth=0.03cm](21.24,11.534744)(20.64,10.934745)
\psframe[linewidth=0.03,dimen=middle](21.64,12.934745)(21.24,11.334744)
\psline[linewidth=0.03cm](19.64,13.134745)(19.24,12.734744)
\psline[linewidth=0.03cm](22.04,12.734744)(21.64,12.334744)
\psline[linewidth=0.03cm](21.64,11.934745)(22.04,12.334744)
\psline[linewidth=0.03cm](21.24,7.1347446)(20.44,7.9347444)
\psline[linewidth=0.03cm](21.24,7.5347443)(20.44,8.334744)
\psline[linewidth=0.03cm](21.24,7.9347444)(20.64,8.534744)
\psframe[linewidth=0.03,dimen=middle](21.64,8.134745)(21.24,6.5347443)
\psline[linewidth=0.03cm](22.04,6.3347445)(21.64,6.7347445)
\psline[linewidth=0.03cm](22.04,6.7347445)(21.64,7.1347446)
\psline[linewidth=0.03cm](21.64,7.5347443)(22.04,7.1347446)
\psframe[linewidth=0.03,dimen=middle](17.24,10.334744)(16.84,9.134745)
\psline[linewidth=0.03cm](16.24,10.134745)(16.84,10.134745)
\psline[linewidth=0.03cm](16.24,9.334744)(16.84,9.334744)
\psline[linewidth=0.03cm](16.24,9.734744)(16.84,9.734744)
\psline[linewidth=0.03cm](21.828957,10.134745)(21.04,10.134745)
\psline[linewidth=0.03cm](21.828957,9.734744)(20.44,9.734744)
\psline[linewidth=0.03cm](21.828957,9.334744)(21.04,9.334744)
\psframe[linewidth=0.03,dimen=middle](22.228958,10.334744)(21.828957,9.134745)
\psline[linewidth=0.03cm](22.634712,10.134745)(22.228958,10.134745)
\psline[linewidth=0.03cm](22.634712,9.334744)(22.228958,9.334744)
\psline[linewidth=0.03cm](22.634712,9.734744)(22.228958,9.734744)
\psline[linewidth=0.03cm](18.64,9.734744)(20.44,9.734744)
\psbezier[linewidth=0.03](18.64,10.134745)(19.84,10.134745)(19.64,10.934745)(20.44,11.534744)
\psbezier[linewidth=0.03](18.64,9.334744)(19.84,9.334744)(19.64,8.734744)(20.44,7.9347444)
\psbezier[linewidth=0.03](20.64,10.934745)(20.24,10.534744)(20.84,10.134745)(21.04,10.134745)
\psbezier[linewidth=0.03](20.64,8.534744)(20.24,8.934745)(20.84,9.334744)(21.04,9.334744)
\psline[linewidth=0.03cm](17.24,-6.4652557)(18.628958,-6.4652557)
\psline[linewidth=0.03cm](17.24,-6.8652554)(18.628958,-6.8652554)
\psline[linewidth=0.03cm](17.24,-7.2652555)(18.628958,-7.2652555)
\psline[linewidth=0.03cm](21.24,-4.2652555)(20.44,-5.0652556)
\psline[linewidth=0.03cm](21.24,-4.6652555)(20.44,-5.4652557)
\psline[linewidth=0.03cm](21.24,-5.0652556)(20.64,-5.6652555)
\psframe[linewidth=0.03,dimen=middle](21.64,-3.6652555)(21.24,-5.2652555)
\psline[linewidth=0.03cm](22.04,-3.4652555)(21.64,-3.8652556)
\psline[linewidth=0.03cm](22.04,-3.8652556)(21.64,-4.2652555)
\psline[linewidth=0.03cm](21.64,-4.6652555)(22.04,-4.2652555)
\psline[linewidth=0.03cm](21.24,-9.465256)(20.44,-8.665256)
\psline[linewidth=0.03cm](21.24,-9.065255)(20.44,-8.265256)
\psline[linewidth=0.03cm](21.24,-8.665256)(20.64,-8.065255)
\psframe[linewidth=0.03,dimen=middle](21.64,-8.465256)(21.24,-10.065255)
\psline[linewidth=0.03cm](22.04,-10.265256)(21.64,-9.865255)
\psline[linewidth=0.03cm](22.04,-9.865255)(21.64,-9.465256)
\psline[linewidth=0.03cm](21.64,-9.065255)(22.04,-9.465256)
\psframe[linewidth=0.03,dimen=middle](17.24,-6.2652555)(16.84,-7.4652557)
\psline[linewidth=0.03cm](16.24,-6.4652557)(16.84,-6.4652557)
\psline[linewidth=0.03cm](16.24,-7.2652555)(16.84,-7.2652555)
\psline[linewidth=0.03cm](16.24,-6.8652554)(16.84,-6.8652554)
\psline[linewidth=0.03cm](21.828957,-6.4652557)(21.04,-6.4652557)
\psline[linewidth=0.03cm](21.828957,-6.8652554)(20.44,-6.8652554)
\psline[linewidth=0.03cm](21.828957,-7.2652555)(21.04,-7.2652555)
\psframe[linewidth=0.03,dimen=middle](22.228956,-6.2652555)(21.828957,-7.4652557)
\psline[linewidth=0.03cm](22.634712,-6.4652557)(22.228956,-6.4652557)
\psline[linewidth=0.03cm](22.634712,-7.2652555)(22.228956,-7.2652555)
\psline[linewidth=0.03cm](22.634712,-6.8652554)(22.228956,-6.8652554)
\psbezier[linewidth=0.03](18.64,-6.4652557)(19.84,-6.4652557)(19.64,-5.6652555)(20.44,-5.0652556)
\psbezier[linewidth=0.03](18.64,-7.2652555)(19.84,-7.2652555)(19.64,-7.8652554)(20.44,-8.665256)
\psbezier[linewidth=0.03](20.64,-5.6652555)(20.24,-6.0652556)(20.24,-6.4652557)(21.04,-6.4652557)
\psbezier[linewidth=0.03](20.64,-8.065255)(20.24,-7.6652555)(20.84,-7.2652555)(21.04,-7.2652555)
\psbezier[linewidth=0.03](18.64,-6.8652554)(19.64,-6.8652554)(19.64,-7.2652555)(20.44,-8.265256)
\psbezier[linewidth=0.03](20.44,-5.4652557)(20.04,-5.8652554)(19.44,-6.8652554)(20.44,-6.8652554)
\usefont{T1}{ptm}{m}{n}
\rput(14.07,-6.8702555){\LARGE $P_{1,3}\quad=$}
\psbezier[linewidth=0.03](20.44,11.134745)(19.64,10.134745)(19.64,9.334744)(20.44,8.334744)
\usefont{T1}{ptm}{m}{n}
\rput(2.54,9.709744){\LARGE $T_{2,2}\quad=$}
\usefont{T1}{ptm}{m}{n}
\rput(14.2,9.749744){\LARGE $T_{1,3}\quad=$}
\psline[linewidth=0.03cm](4.84,3.5347445)(5.64,2.7347443)
\psline[linewidth=0.03cm](4.84,3.1347444)(5.64,2.3347445)
\psline[linewidth=0.03cm](4.84,2.7347443)(5.44,2.1347444)
\psframe[linewidth=0.03,dimen=middle](4.84,4.1347446)(4.44,2.5347445)
\psline[linewidth=0.03cm](4.04,4.3347445)(4.44,3.9347444)
\psline[linewidth=0.03cm](4.04,3.9347444)(4.44,3.5347445)
\psline[linewidth=0.03cm](4.44,3.1347444)(4.04,3.5347445)
\psline[linewidth=0.03cm](8.64,3.5347445)(7.84,2.7347443)
\psline[linewidth=0.03cm](8.64,3.1347444)(7.84,2.3347445)
\psline[linewidth=0.03cm](8.64,2.7347443)(8.04,2.1347444)
\psframe[linewidth=0.03,dimen=middle](9.04,4.1347446)(8.64,2.5347445)
\psline[linewidth=0.03cm](9.44,4.3347445)(9.04,3.9347444)
\psline[linewidth=0.03cm](9.44,3.9347444)(9.04,3.5347445)
\psline[linewidth=0.03cm](9.04,3.1347444)(9.44,3.5347445)
\psline[linewidth=0.03cm](4.84,-0.4652556)(5.64,0.33474442)
\psline[linewidth=0.03cm](4.84,-0.065255575)(5.64,0.7347444)
\psline[linewidth=0.03cm](4.84,0.33474442)(5.44,0.9347444)
\psframe[linewidth=0.03,dimen=middle](4.84,0.53474444)(4.44,-1.0652555)
\psline[linewidth=0.03cm](4.04,-1.2652556)(4.44,-0.8652556)
\psline[linewidth=0.03cm](4.04,-0.8652556)(4.44,-0.4652556)
\psline[linewidth=0.03cm](4.44,-0.065255575)(4.04,-0.4652556)
\psline[linewidth=0.03cm](8.64,-0.4652556)(7.84,0.33474442)
\psline[linewidth=0.03cm](8.64,-0.065255575)(7.84,0.7347444)
\psline[linewidth=0.03cm](8.64,0.33474442)(8.04,0.9347444)
\psframe[linewidth=0.03,dimen=middle](9.04,0.53474444)(8.64,-1.0652555)
\psline[linewidth=0.03cm](9.44,-1.2652556)(9.04,-0.8652556)
\psline[linewidth=0.03cm](9.44,-0.8652556)(9.04,-0.4652556)
\psline[linewidth=0.03cm](9.04,-0.065255575)(9.44,-0.4652556)
\psbezier[linewidth=0.03](5.44,2.1347444)(5.84,1.7347444)(5.84,1.3347445)(5.44,0.9347444)
\psbezier[linewidth=0.03](8.04,2.1347444)(7.64,1.7347444)(7.64,1.3347445)(8.04,0.9347444)
\psbezier[linewidth=0.03](5.64,2.7347443)(6.24,2.1347444)(7.24,2.1347444)(7.84,2.7347443)
\psbezier[linewidth=0.03](5.64,0.33474442)(6.24,0.9347444)(7.24,0.9347444)(7.84,0.33474442)
\psbezier[linewidth=0.03](5.64,2.3347445)(6.24,1.7347444)(6.24,1.3347445)(5.64,0.7347444)
\psbezier[linewidth=0.03](7.84,2.3347445)(7.24,1.7347444)(7.24,1.3347445)(7.84,0.7347444)
\usefont{T1}{ptm}{m}{n}
\rput(2.36,1.4297445){\LARGE $P_{2,2}^1\quad=$}
\psline[linewidth=0.03cm](17.44,3.5347445)(18.24,2.7347443)
\psline[linewidth=0.03cm](17.44,3.1347444)(18.24,2.3347445)
\psline[linewidth=0.03cm](17.44,2.7347443)(18.04,2.1347444)
\psframe[linewidth=0.03,dimen=middle](17.44,4.1347446)(17.04,2.5347445)
\psline[linewidth=0.03cm](16.64,4.3347445)(17.04,3.9347444)
\psline[linewidth=0.03cm](16.64,3.9347444)(17.04,3.5347445)
\psline[linewidth=0.03cm](17.04,3.1347444)(16.64,3.5347445)
\psline[linewidth=0.03cm](21.24,3.5347445)(20.44,2.7347443)
\psline[linewidth=0.03cm](21.24,3.1347444)(20.44,2.3347445)
\psline[linewidth=0.03cm](21.24,2.7347443)(20.64,2.1347444)
\psframe[linewidth=0.03,dimen=middle](21.64,4.1347446)(21.24,2.5347445)
\psline[linewidth=0.03cm](22.04,4.3347445)(21.64,3.9347444)
\psline[linewidth=0.03cm](22.04,3.9347444)(21.64,3.5347445)
\psline[linewidth=0.03cm](21.64,3.1347444)(22.04,3.5347445)
\psline[linewidth=0.03cm](17.44,-0.4652556)(18.24,0.33474442)
\psline[linewidth=0.03cm](17.44,-0.065255575)(18.24,0.7347444)
\psline[linewidth=0.03cm](17.44,0.33474442)(18.04,0.9347444)
\psframe[linewidth=0.03,dimen=middle](17.44,0.53474444)(17.04,-1.0652555)
\psline[linewidth=0.03cm](16.64,-1.2652556)(17.04,-0.8652556)
\psline[linewidth=0.03cm](16.64,-0.8652556)(17.04,-0.4652556)
\psline[linewidth=0.03cm](17.04,-0.065255575)(16.64,-0.4652556)
\psline[linewidth=0.03cm](21.24,-0.4652556)(20.44,0.33474442)
\psline[linewidth=0.03cm](21.24,-0.065255575)(20.44,0.7347444)
\psline[linewidth=0.03cm](21.24,0.33474442)(20.64,0.9347444)
\psframe[linewidth=0.03,dimen=middle](21.64,0.53474444)(21.24,-1.0652555)
\psline[linewidth=0.03cm](22.04,-1.2652556)(21.64,-0.8652556)
\psline[linewidth=0.03cm](22.04,-0.8652556)(21.64,-0.4652556)
\psline[linewidth=0.03cm](21.64,-0.065255575)(22.04,-0.4652556)
\psbezier[linewidth=0.03](18.04,2.1347444)(18.44,1.7347444)(18.44,1.3347445)(18.04,0.9347444)
\psbezier[linewidth=0.03](20.64,2.1347444)(20.24,1.7347444)(20.24,1.3347445)(20.64,0.9347444)
\psbezier[linewidth=0.03](18.24,2.7347443)(18.84,2.1347444)(19.84,2.1347444)(20.44,2.7347443)
\psbezier[linewidth=0.03](18.24,0.33474442)(18.84,0.9347444)(19.84,0.9347444)(20.44,0.33474442)
\usefont{T1}{ptm}{m}{n}
\rput(14.8,1.5097444){\LARGE $P_{2,2}^2\quad=$}
\psbezier[linewidth=0.03](18.24,0.7347444)(18.84,1.3347445)(19.84,1.3347445)(20.44,0.7347444)
\psbezier[linewidth=0.03](18.24,2.3347445)(18.84,1.7347444)(19.84,1.7347444)(20.44,2.3347445)
\psbezier[linewidth=0.03](5.64,10.534744)(6.44,9.734744)(7.04,9.734744)(7.84,8.934745)
\psbezier[linewidth=0.03](7.84,10.534744)(7.04,9.734744)(6.44,9.734744)(5.64,8.934745)
\psline[linewidth=0.03cm](4.84,-5.0652556)(5.64,-5.8652554)
\psline[linewidth=0.03cm](4.84,-5.4652557)(5.64,-6.2652555)
\psline[linewidth=0.03cm](4.84,-5.8652554)(5.44,-6.4652557)
\psframe[linewidth=0.03,dimen=middle](4.84,-4.4652557)(4.44,-6.0652556)
\psline[linewidth=0.03cm](4.04,-4.2652555)(4.44,-4.6652555)
\psline[linewidth=0.03cm](4.04,-4.6652555)(4.44,-5.0652556)
\psline[linewidth=0.03cm](4.44,-5.4652557)(4.04,-5.0652556)
\psline[linewidth=0.03cm](8.64,-5.0652556)(7.84,-5.8652554)
\psline[linewidth=0.03cm](8.64,-5.4652557)(7.84,-6.2652555)
\psline[linewidth=0.03cm](8.64,-5.8652554)(8.04,-6.4652557)
\psframe[linewidth=0.03,dimen=middle](9.04,-4.4652557)(8.64,-6.0652556)
\psline[linewidth=0.03cm](9.44,-4.2652555)(9.04,-4.6652555)
\psline[linewidth=0.03cm](9.44,-4.6652555)(9.04,-5.0652556)
\psline[linewidth=0.03cm](9.04,-5.4652557)(9.44,-5.0652556)
\psline[linewidth=0.03cm](4.84,-9.065255)(5.64,-8.265256)
\psline[linewidth=0.03cm](4.84,-8.665256)(5.64,-7.8652554)
\psline[linewidth=0.03cm](4.84,-8.265256)(5.44,-7.6652555)
\psframe[linewidth=0.03,dimen=middle](4.84,-8.065255)(4.44,-9.665256)
\psline[linewidth=0.03cm](8.64,-9.065255)(7.84,-8.265256)
\psline[linewidth=0.03cm](8.64,-8.665256)(7.84,-7.8652554)
\psline[linewidth=0.03cm](8.64,-8.265256)(8.04,-7.6652555)
\psframe[linewidth=0.03,dimen=middle](9.04,-8.065255)(8.64,-9.665256)
\psbezier[linewidth=0.03](5.44,-6.4652557)(5.84,-6.8652554)(5.84,-7.2652555)(5.44,-7.6652555)
\psbezier[linewidth=0.03](8.04,-6.4652557)(7.64,-6.8652554)(7.64,-7.2652555)(8.04,-7.6652555)
\psbezier[linewidth=0.03](5.64,-5.8652554)(6.24,-6.4652557)(7.24,-6.4652557)(7.84,-5.8652554)
\psbezier[linewidth=0.03](5.64,-8.265256)(6.24,-7.6652555)(7.24,-7.6652555)(7.84,-8.265256)
\psbezier[linewidth=0.03](5.64,-6.2652555)(6.24,-6.8652554)(6.24,-7.2652555)(5.64,-7.8652554)
\psbezier[linewidth=0.03](7.84,-6.2652555)(7.24,-6.8652554)(7.24,-7.2652555)(7.84,-7.8652554)
\psbezier[linewidth=0.03](9.0,-8.450255)(9.434559,-8.883589)(12.186772,-10.761367)(9.869119,-11.050256)
\psbezier[linewidth=0.03](9.0,-8.850256)(9.434559,-9.250256)(11.6,-10.650255)(9.579413,-10.850256)
\psbezier[linewidth=0.03](9.0,-9.250256)(9.6,-9.850256)(11.0,-10.650255)(9.2,-10.650255)
\psbezier[linewidth=0.03](9.2,-10.650255)(7.56,-10.650255)(6.4,-11.450255)(3.8,-12.450255)
\psbezier[linewidth=0.03](9.6,-10.850256)(7.52,-10.810255)(4.8,-12.450255)(4.0,-12.650255)
\psbezier[linewidth=0.03](9.88,-11.050255)(7.04,-11.250255)(5.8,-12.250256)(4.2,-12.850256)
\psbezier[linewidth=0.03](4.3867717,-8.450255)(3.952212,-8.883589)(1.2,-10.761367)(3.5176523,-11.050256)
\psbezier[linewidth=0.03](4.3867717,-8.850256)(3.952212,-9.250256)(1.7867718,-10.650255)(3.8073587,-10.850256)
\psbezier[linewidth=0.03](4.3867717,-9.250256)(3.7867718,-9.850256)(2.3867717,-10.650255)(4.186772,-10.650255)
\psbezier[linewidth=0.03](4.16,-10.650255)(5.069091,-10.670255)(6.0,-11.050256)(6.4,-11.250256)
\psbezier[linewidth=0.03](3.8,-10.850256)(4.7090907,-10.850256)(5.56,-11.030255)(6.12,-11.410255)
\psbezier[linewidth=0.03](3.52,-11.050256)(4.12,-11.050256)(5.06,-11.130256)(5.8,-11.550256)
\psbezier[linewidth=0.03](7.2,-11.790256)(8.04,-12.190255)(8.92,-12.550256)(9.32,-12.770256)
\psbezier[linewidth=0.03](6.876964,-11.910255)(7.6155357,-12.298983)(8.778786,-12.765454)(9.185,-12.979254)
\psbezier[linewidth=0.03](6.6,-12.046309)(7.3385715,-12.435037)(8.612608,-12.959818)(8.963429,-13.134745)
\usefont{T1}{ptm}{m}{n}
\rput(2.56,-7.1702557){\LARGE $P_{2,2}^3\quad=$}
\end{pspicture} 
}
\end{center}
\caption{The vertex operators. In the pillow case, the upper index labels the three different channels.}
\label{operators}
\end{figure}

Now, the `tetrahedral' and `pillow' vertices and the covariance $C_{\phi}$ can be combined into operators

\bea
T_{2,2}:\cH^{(\La)}_{0}\otimes\cH^{(\La)}_{0} & \longrightarrow & \cH^{(\La)}_{0}\otimes\cH^{(\La)}_{0}\nonumber\\
T_{1,3}:\cH^{(\La)}_{0} & \longrightarrow & \cH^{(\La)}_{0}\otimes\cH^{(\La)}_{0}\otimes\cH^{(\La)}_{0}\nonumber\\
P_{2,2}^{\alpha}:\cH^{(\La)}_{0}\otimes\cH^{(\La)}_{0} & \longrightarrow & \cH^{(\La)}_{0}\otimes\cH^{(\La)}_{0}\nonumber\\
P_{1,3}:\cH^{(\La)}_{0} & \longrightarrow & \cH^{(\La)}_{0}\otimes\cH^{(\La)}_{0}\otimes\cH^{(\La)}_{0},
\eea
where $\alpha=1,2,3$ is a channel index, see Fig. \ref{operators}, in such a way that any BFL Feynman amplitude is obtained as the trace of a certain product of these operators and their adjoints. They satisfy the following norm bounds:

\begin{lemma}\label{vertexbounds}
\bea  &&\Vert T_{2,2} \Vert \le K ,\quad \Vert T_{1,3} \Vert   \le K \Lambda^{3/2}\nonumber\\
 &&\Vert P_{2,2}^1 \Vert \le K \Lambda^3 ,\quad\Vert P_{2,2}^2 \Vert   \le K,  \nonumber\\&&\Vert P_{2,2}^3\Vert\le K,\quad \Vert P_{1,3} \Vert
 \le K \Lambda^{3}.
\eea
\end{lemma}

% Generated with LaTeXDraw 2.0.2
% Mon Jun 29 13:04:40 CEST 2009
% \usepackage[usenames,dvipsnames]{pstricks}
% \usepackage{epsfig}
% \usepackage{pst-grad} % For gradients
% \usepackage{pst-plot} % For axes

\prf
To evaluate these norms, we can use the formula
\be   \Vert H \Vert = \lim_{n \to \infty}  \biggl( {\rm Tr }  [ HH^t ]^n  \biggr)^{1/2n}.
\ee

\begin{figure}[h]
\begin{center}
\scalebox{0.7} % Change this value to rescale the drawing.
{
\begin{pspicture}(0,-2.515)(11.915,2.515)
\psdots[dotsize=0.08](3.3,1.1)
\psdots[dotsize=0.08](4.5,1.1)
\psbezier[linewidth=0.03](3.3,1.1)(3.7,0.7)(4.1,0.7)(4.5,1.1)
\psbezier[linewidth=0.03](3.3,1.1)(3.7,1.5)(4.1,1.5)(4.5,1.1)
\psbezier[linewidth=0.03](8.1,1.1)(8.5,0.7)(8.9,0.7)(9.3,1.1)
\psbezier[linewidth=0.03](8.1,1.1)(8.5,1.5)(8.9,1.5)(9.3,1.1)
\psbezier[linewidth=0.03](6.9,1.1)(7.3,0.7)(7.7,0.7)(8.1,1.1)
\psbezier[linewidth=0.03](6.9,1.1)(7.3,1.5)(7.7,1.5)(8.1,1.1)
\psbezier[linewidth=0.03](5.7,1.1)(6.1,0.7)(6.5,0.7)(6.9,1.1)
\psbezier[linewidth=0.03](5.7,1.1)(6.1,1.5)(6.5,1.5)(6.9,1.1)
\psdots[dotsize=0.08](4.5,1.1)
\psbezier[linewidth=0.03](4.5,1.1)(4.9,0.7)(5.3,0.7)(5.7,1.1)
\psbezier[linewidth=0.03](4.5,1.1)(4.9,1.5)(5.3,1.5)(5.7,1.1)
\psbezier[linewidth=0.03](9.3,1.1)(9.7,0.7)(10.1,0.7)(10.5,1.1)
\psbezier[linewidth=0.03](9.3,1.1)(9.7,1.5)(10.1,1.5)(10.5,1.1)
\psdots[dotsize=0.08](5.7,1.1)
\psdots[dotsize=0.08](6.9,1.1)
\psdots[dotsize=0.08](8.1,1.1)
\psdots[dotsize=0.08](9.3,1.1)
\psdots[dotsize=0.08](10.5,1.1)
\psbezier[linewidth=0.03](3.3,1.1)(2.3,2.1)(5.3,2.5)(6.9,2.5)
\psbezier[linewidth=0.03](10.5,1.1)(11.5,2.1)(8.5,2.5)(6.9,2.5)
\psbezier[linewidth=0.03](3.3,1.1)(2.3,0.1)(5.3,-0.3)(6.9,-0.3)
\psbezier[linewidth=0.03](10.5,1.1)(11.5,0.1)(8.5,-0.3)(6.9,-0.3)
\psline[linewidth=0.03cm](3.3,-1.7)(4.5,-1.7)
\psbezier[linewidth=0.03](3.5,-1.7)(3.7,-2.1)(4.1,-2.1)(4.3,-1.7)
\psbezier[linewidth=0.03](3.5,-1.7)(3.7,-1.3)(4.1,-1.3)(4.3,-1.7)
\psline[linewidth=0.03cm](4.3,-1.7)(5.5,-1.7)
\psbezier[linewidth=0.03](4.5,-1.7)(4.7,-2.1)(5.1,-2.1)(5.3,-1.7)
\psbezier[linewidth=0.03](4.5,-1.7)(4.7,-1.3)(5.1,-1.3)(5.3,-1.7)
\psline[linewidth=0.03cm](5.3,-1.7)(6.5,-1.7)
\psbezier[linewidth=0.03](5.5,-1.7)(5.7,-2.1)(6.1,-2.1)(6.3,-1.7)
\psbezier[linewidth=0.03](5.5,-1.7)(5.7,-1.3)(6.1,-1.3)(6.3,-1.7)
\psline[linewidth=0.03cm](7.3,-1.7)(8.5,-1.7)
\psbezier[linewidth=0.03](7.5,-1.7)(7.7,-2.1)(8.1,-2.1)(8.3,-1.7)
\psbezier[linewidth=0.03](7.5,-1.7)(7.7,-1.3)(8.1,-1.3)(8.3,-1.7)
\psline[linewidth=0.03cm](6.3,-1.7)(7.5,-1.7)
\psbezier[linewidth=0.03](6.5,-1.7)(6.7,-2.1)(7.1,-2.1)(7.3,-1.7)
\psbezier[linewidth=0.03](6.5,-1.7)(6.7,-1.3)(7.1,-1.3)(7.3,-1.7)
\psline[linewidth=0.03cm](8.3,-1.7)(9.5,-1.7)
\psbezier[linewidth=0.03](8.5,-1.7)(8.7,-2.1)(9.1,-2.1)(9.3,-1.7)
\psbezier[linewidth=0.03](8.5,-1.7)(8.7,-1.3)(9.1,-1.3)(9.3,-1.7)
\psline[linewidth=0.03cm](9.3,-1.7)(10.5,-1.7)
\psbezier[linewidth=0.03](9.5,-1.7)(9.7,-2.1)(10.1,-2.1)(10.3,-1.7)
\psbezier[linewidth=0.03](9.5,-1.7)(9.7,-1.3)(10.1,-1.3)(10.3,-1.7)
\psbezier[linewidth=0.03](3.3,-1.7)(1.9,-1.7)(3.5,-2.5)(6.9,-2.5)
\psbezier[linewidth=0.03](10.5,-1.7)(11.9,-1.7)(10.3,-2.5)(6.9,-2.5)
\usefont{T1}{ptm}{m}{n}
\rput(1.41,1.105){\Large $G_n\quad=$}
\usefont{T1}{ptm}{m}{n}
\rput(1.33,-1.695){\Large $G'_n\quad=$}
\end{pspicture} 
}
\end{center}
\caption{The chains of graphs $G_{n}$ and $G'_{n}$, each with $n$ vertices.}
\label{sequences}
\end{figure}
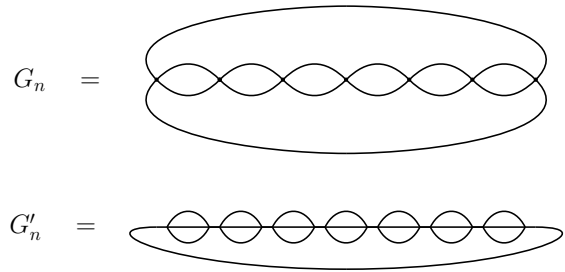

In the Boulatov model, this leads respectively to the computation of the chains $G_n$ 
for $T_{2,2}$ and $G'_n$ for $T_{1,3}$ as in Fig. \ref{sequences}.

% Generated with LaTeXDraw 2.0.2
% Fri Jun 26 17:20:53 CEST 2009
% \usepackage[usenames,dvipsnames]{pstricks}
% \usepackage{epsfig}
% \usepackage{pst-grad} % For gradients
% \usepackage{pst-plot} % For axes

These chain are very simple and can be evaluated easily directly; they are also 
particular cases of `type I' graphs, whose amplitudes were evaluated in \cite{Freidel:2009hd}. In the Boulatov case, one finds $\cA^B_{G_n}  \simeq  K^n $, $\cA^B_{G'_n}  \simeq  K^n \Lambda^{6+3n/2}$  (the factor
$6$ comes from the final trace).

In the BFL case, the dominant graphs are the chains
$G_n$ in the channel $P_{2,2}^{1}$ and $G'_{n}$ with $P_{1,3}$, for which $\cA^{BFL,1}_{G_n}  \simeq  K^n \Lambda^{6+3n}$. The other chains are subdominant, and yield the results given in (\ref{vertexbounds}).\qed

Remarking that a two-point graph is a $\cH^{(\La)}_{0} \to \cH^{(\La)}_{0}$ operator, this shows that for
a two-point graph without tadpoles, the norm of the corresponding operator is bounded by 
$K^n \La^{3n/2}$ and  $K^n \La^{3n}$ in the Boulatov and BFL cases respectively.
Indeed, the existence of an exhausting sequence of cuts precisely allows to write the operator
for that two-point graph as the composition of $n$ operators, each made of a tensor product of
certain number of times the identity times a single vertex operator. Applying the bounds of (\ref{vertexbounds}),
we conclude that
Theorems \ref{pertB} and \ref{pertBFL} must hold for any graphs without 
generalized tadpoles: a vacuum graph is the trace of a two-point subgraph,
and that  trace costs at most $\Lambda^{6}$.

But a two-point generalized tadpole graph $G$ with sub-generalized tadpoles is again
a two-point subgraph without tadpole joined to a $T_{2,2}$ or $P _{2,2}$ operator, so its norm can be bounded
again in the same way by an easy induction. This completes the proof of Theorems \ref{pertB} and \ref{pertBFL}. Finally the examples of graphs $G_n$ and $G'_n$ prove Theorem \ref{satur}.

The Theorem  \ref{pertBFL}, which express the perturbative scaling behaviour of the BFL model, can be upgraded to the constructive level -- which is what we now turn to.

\section{Constructive field theory via cactus expansions}\label{constructive}

The historic method in Bosonic constructive field theory is to first introduce a discretization of space-time, e.g.
through a lattice of cubes, then test the couplings between the corresponding functional integrals restricted to these cubes.
Since the interaction is local these couplings occur only through propagators of the theory joining different cubes.
This expansion, called the \emph{cluster expansion}, results in the theory being written as a polymer gas with hardcore constraints. For that gas to be dilute at small coupling, the normalization of the free functional integrals must be factored out.
Finally the connected functions are computed by expanding away the hardcore constraint through a so-called
Mayer expansion \cite{GJ,Riv,Br,AR}. This last step implies a replica trick in some way or another.
Both steps are essential; organizing functional integrals around decoupled degrees of freedom is the essence
of perturbation theory and computing connected functions, or logarithms of the partition function
is also a key step in quantum field theory. The renormalization group can be seen as essentially the iteration
of these two steps, namely functional integration over a fluctuation field followed by the computation 
of a logarithm in order to find an effective action. 

However this historic constructive method, at least in its standard formulation, is 
unsuited for non-local field theories. The discretization of the space $\su^3$ into 
hypercubes does not make sense, and the non-local interaction won't factorize anyway. 
A similar difficulty arose in the Grosse-Wulkenhaar non-commutative field theory \cite{Grosse:2004yu},
which is non-local in direct space or a matrix model in the matrix base. It was in order to solve this problem
that another constructive method was recently invented, the cactus (or `loop vertex') expansion \cite{Rivasseau:2007fr}. This method gives correct estimates for matrix-like models, and also (as shown in this paper)
can be adapted to more general models of the tensorial type such as the (regularized) Boulatov model.

The cactus expansion can be thought 
as a completely explicit convergent reorganization of standard perturbation theory.
Its beauty is that, 
being closer to Feynman graphs than the cluster and Mayer expansions, it is better suited to treat more exotic situations
such as the non-local interactions of group field theory.
Another advantage of the cactus expansion is that in some sense it performs the two essential steps of constructive 
theory (the cluster and Mayer expansions) \emph{at once}: being expressed as a sum of graph amplitudes factorizing over connected components, the computation of the connected functions (such as the free energy) is straightforward -- one simply restricts the sum to connected graphs. 

We illustrate below this cactus expansion on a simple toy model, namely the $\phi^4$ field theory in $0$ dimension.\footnote{For a historical perspective on constructive methods applied to this $0$-dimensional model, see \cite{CFT0d}.}

\subsection{Four ingredients for a cactus expansion}
The recipe to perform the cactus expansion of a free function (or any connected correlation function) has four ingredients: (1) an intermediate field, (2) a resolvent bound, (3) a replica trick and (4) a forest formula.

\begin{enumerate}
\item The intermediate field representation is a well-known trick to represent a quartic
interaction $\int\phi^4$ in terms of a cubic one $\int\phi\sigma\phi$:
\be
e^{-\lambda\int\phi^4/8}=\int d\nu(\sigma)\ e^{-\f{1}{2}i\sqrt{\lambda}\int\phi \sigma\phi},
\ee
where $\sigma$ is a field with Gaussian ultralocal measure $d\nu$ (which in zero dimension
would simply be the Gaussian measure over $\mathbb{R}$). Performing the Gaussian integration over $\phi$, 
one can then express the original path integral $Z(\lambda)$ in terms of $\sigma$ only, for instance in zero dimension as
\be
Z(\lambda)= \int d\nu(\sigma) \int d\mu(\phi)\ e^{-\f{1}{2}i\sqrt{\lambda}\int \phi \sigma\phi}=\int d\nu(\sigma) \det(1+i\sqrt{\lambda}\sigma)^{-1/2}=\int d\nu(\sigma) e^{-\f{1}{2} \text{Tr}\ \text{Log}(1+i\sqrt{\lambda}\sigma)}.
\ee 
In more than zero dimension, a propagator $C^{1/2}$ would typically sandwich the $\sigma$ field on both sides. 
This process generate a `loop vertex' \cite{Rivasseau:2007fr} for the intermediate field $\sigma$ of the form $\text{Tr}\ \text{Log}(1+i\sqrt{\lambda}\sigma)$ where $\text{Tr}$ stands for the operator trace. This trick can be generalized to any correlation functions and to more complicated models.
\item
The replica trick relies on the properties of degenerate Gaussian measures. Let  $d\nu(\sigma)$ denote the standard Gaussian measure on $\mathbb{R}$, and $V\in L^n(\mathbb{R})$. Now, `replicate' $n$ times the variable $\sigma$, and consider the degenerate (normalized, centered) Gaussian measure $d\nu_n$ on $\mathbb{R}^n$ with covariance $C_{ij}=\la\sigma_i\sigma_j\ra=1$. The replica trick is the statement that\footnote{Let us emphasize that this is not a form of Fubini's theorem, which expresses an $n$-dimensional integral as a product of $n$ integrals. Here, we replace {\it one} $1$-dimensional integral by an $n$-dimensional one, with $(n-1)$ delta functions.}
\be
\int d\nu(\sigma)\ V(\sigma)^n=\int d\nu_n(\sigma_1,\dots,\sigma_n)\ \prod_{v=1}^nV(\sigma_v).
\ee
A general discussion on degenerate Gaussian measures is in Appendix A.
\item
Consider a smooth function $H$ of $\f{n(n-1)}{2}$ variables $\boldsymbol{h}=(h_{l})$, living on the lines $l$ of the complete graph over $n$ vertices. The so-called Brydges-Kennedy Taylor forest formula \cite{Abdesselam:1994ap,BK} is a Taylor interpolation of $H$ with integral remainders indexed by labeled forests over $n$ vertices (see Appendix B):

\be\label{forest}
H(\boldsymbol{1})=\sum_{F\in\mathcal{F}_n}\left(\prod_{l\in F}\int_0^1dh_l\right)\left(\prod_{l\in F}\f{\partial}{\partial h_l}\right)H(\boldsymbol{h}^F).
\ee
In this expression, $\cF_n$ denotes the set of forests over $n$ vertices, the products are over lines $l$ of each forest $F$, and $\boldsymbol{h}^F$ is the $\f{n(n-1)}{2}$-uple defined by $h^F_l:=\min_ph_p$, where $p$ runs over the unique path in $F$ connecting the source and target vertices of $l$. (If they are not connected by $F$, then $h^F_l:=0$.) 

One can easily check that for $n=2$, this is nothing but the fundamental theorem of calculus: $H(1)=H(0)+\int_{0}^1dh\ H'(h)$. For higher values of $n$, on the other hand, the outcome of (\ref{forest}) is genuinely non-trivial, as the case $n=3$ already demonstrates (Fig. \ref{forests}):

\begin{eqnarray}\label{exforest}
H(1,1,1)&=&H(0,0,0)+\int_0^1 dh_1\ \partial_1H(h_1,0,0)+\int_0^1 dh_2\ \partial_2H(0,h_2,0)+\int_0^1 dh_3\ \partial_3H(0,0,h_3)\nonumber\\
&& + \int_0^1dh_1\int_0^1dh_2\ \partial^2_{12}H(h_1,h_2,\min(h_1,h_2))+\int_0^1dh_1\int_0^1dh_3\ \partial^2_{13}H(h_1,\min(h_1,h_3),h_3)
\nonumber\\ 
&& + \int_0^1dh_2\int_0^1dh_3\ \partial^2_{23}H(\min(h_2,h_3),h_2,h_3).
\end{eqnarray}

% Generated with LaTeXDraw 2.0.2
% Fri Mar 06 16:37:00 CET 2009
% \usepackage[usenames,dvipsnames]{pstricks}
% \usepackage{epsfig}
% \usepackage{pst-grad} % For gradients
% \usepackage{pst-plot} % For axes
\begin{figure}[h]
\begin{center}
\scalebox{1} % Change this value to rescale the drawing.
{
\begin{pspicture}(0,-1.37)(10.56,1.39)
\psdots[dotsize=0.12](0.08,-1.29)
\psdots[dotsize=0.12](0.48,-0.69)
\psdots[dotsize=0.12](0.88,-1.29)
\psdots[dotsize=0.12](1.68,-1.29)
\psdots[dotsize=0.12](2.08,-0.69)
\psdots[dotsize=0.12](2.48,-1.29)
\psline[linewidth=0.04cm](2.08,-0.69)(1.68,-1.29)
\psdots[dotsize=0.12](3.28,-1.29)
\psdots[dotsize=0.12](3.68,-0.69)
\psdots[dotsize=0.12](4.08,-1.29)
\psdots[dotsize=0.12](4.88,-1.29)
\psdots[dotsize=0.12](5.28,-0.69)
\psdots[dotsize=0.12](5.68,-1.29)
\psdots[dotsize=0.12](6.48,-1.29)
\psdots[dotsize=0.12](6.88,-0.69)
\psdots[dotsize=0.12](7.28,-1.29)
\psdots[dotsize=0.12](8.08,-1.29)
\psdots[dotsize=0.12](8.88,-1.29)
\psdots[dotsize=0.12](8.48,-0.69)
\psdots[dotsize=0.12](9.68,-1.29)
\psdots[dotsize=0.12](10.08,-0.69)
\psdots[dotsize=0.12](10.48,-1.29)
\psline[linewidth=0.04cm](3.68,-0.69)(4.08,-1.29)
\psline[linewidth=0.04cm](4.88,-1.29)(5.68,-1.29)
\psline[linewidth=0.04cm](6.48,-1.29)(6.88,-0.69)
\psline[linewidth=0.04cm](6.88,-0.69)(7.28,-1.29)
\psline[linewidth=0.04cm](8.48,-0.69)(8.08,-1.29)
\psline[linewidth=0.04cm](8.08,-1.29)(8.88,-1.29)
\psline[linewidth=0.04cm](10.08,-0.69)(10.48,-1.29)
\psline[linewidth=0.04cm](9.68,-1.29)(10.48,-1.29)
\psdots[dotsize=0.12](4.68,0.31)
\psdots[dotsize=0.12](5.88,0.31)
\psdots[dotsize=0.12](5.28,1.31)
\psline[linewidth=0.04cm](5.28,1.31)(5.88,0.31)
\psline[linewidth=0.04cm](5.88,0.31)(4.68,0.31)
\psline[linewidth=0.04cm](4.68,0.31)(5.28,1.31)
\usefont{T1}{ptm}{m}{n}
\rput(4.73,0.855){\small 1}
\usefont{T1}{ptm}{m}{n}
\rput(5.8,0.875){\small 2}
\usefont{T1}{ptm}{m}{n}
\rput(5.27,0.055){\small 3}
\end{pspicture} 
}
\end{center}
\caption{The complete graph over 3 vertices, and its 7 forests, matching the $7$ terms in (\ref{exforest}).}
\label{forests}
\end{figure}
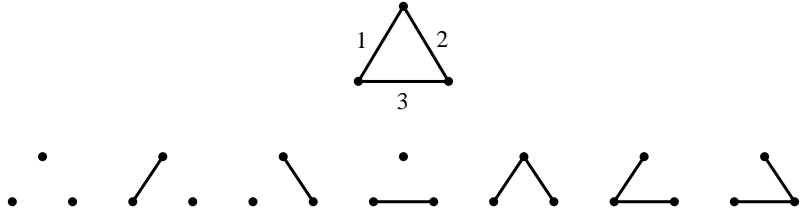

\item
If $\lambda$ is a complex number with positive real part, and $\sigma$ a real number, one has
\be\label{ineq}
\vert1+i\sqrt{\lambda}\sigma\vert^{-1}\leq\sqrt{2}.
\ee
This inequality easily extends to the case where $\sigma$ is replaced by a Hermitian matrix $\Sigma$. Indeed, since the spectral radius of a normal matrix (and the resolvent $(1+i\sqrt{\lambda}\Sigma)^{-1}$ is normal if $\Sigma$ is Hermitian) equals its operator norm, taking the supremum over the spectrum of $\Sigma$ in (\ref{ineq}) gives the `resolvent bound'

\be
\Vert(1+i\sqrt{\lambda}\Sigma)^{-1}\Vert\leq\sqrt{2},
\ee
in which $\Vert\cdot\Vert$ denotes the operator norm. As we shall see in the following, this bound is key to the summability of the cactus expansion.

\end{enumerate}

%%%%%%%%%%
\subsection{A toy example}

As a toy example, consider the $0$-dimensional $\phi^4$ partition function, written in terms of the intermediate `field' $\sigma$,
\be\label{toy}
Z(\lambda):=\int_{\mathbb{R}} d\mu(\phi)\ e^{-\lambda\phi^4/8}=\int_{\mathbb{R}} d\nu(\sigma)\ e^{V_{\lambda}(\sigma)},
\ee
where $V_{\lambda}(\sigma)=-\f{1}{2}\textrm{Log}(1+i\sqrt{\lambda}\sigma)$ and $d\nu(\sigma)$ is the standard Gaussian measure over the real line. The function $F(\lambda):=\log Z(\lambda)$ thus defined is analytic in the cut plane $\mathbb{C}\setminus\mathbb{R}_-$, and hence admits, at best, a Borel expansion about $\lambda=0$. Using the four ingredients presented above, we now show that this is indeed the case.

Expanding the exponential in powers and swapping integration and summation\footnote{Of course, it is precisely such an interchange between integration and summation that yields the divergent perturbative series. Note that here, however, the process is licit because $\int d\mu(\sigma)\ e^{1/2\vert\mathrm{Log}(1+i\sqrt{\lambda}\sigma)\vert}<\infty$, and so Lebesgue's dominated convergence theorem applies.} and applying the replica trick to the order-$n$ term yields
\be\label{sum}
Z(\lambda)=\sum_{n=0}^{\infty}\f{1}{n!}\int d\nu_n(\sigma_1,\dots,\sigma_n)\ \prod_{v=1}^nV_{\lambda}(\sigma_v).
\ee
Next, consider the modified covariance $C^{\boldsymbol{h}}$, parametrized by an $\f{n(n-1)}{2}$-uple $\boldsymbol{h}$, defined by $C^{\boldsymbol{h}}_{ii}:=1$ and $C^{\boldsymbol{h}}_l:=h_l$, where $l=\{ij\}$ ($i\neq j$), and let $d\nu^{\boldsymbol{h}}_n(\sigma_1,\dots,\sigma_n)$ be the corresponding (normalized, centered) Gaussian measure. Applying the forest formula to the function

\be
H(\boldsymbol{h}):=\int d\nu_n^{\boldsymbol{h}}(\sigma_1,\dots,\sigma_n)\ \prod_{v=1}^nV_{\lambda}(\sigma_v),
\ee
we obtain

\be\label{toy loop vertex}
Z(\lambda)=\sum_{n=0}^{\infty}\f{1}{n!}\sum_{F\in\mathcal{F}_n}\left(\prod_{l\in F}\int_0^1dh_l\right)\left(\prod_{l\in F}\f{\partial}{\partial h_l}\right)\int d\nu^{\boldsymbol{h^F}}_n(\sigma_1,\dots,\sigma_n)\prod_{v=1}^nV_{\lambda}(\sigma_v).
\ee

As announced, the summand factorizes along connected components of each forest, yielding a similar formula for the free energy $F(\lambda)$, except for the fact that only connected forests -- trees -- contribute:
\be\label{toytrees}
F(\lambda)=\sum_{n=1}^{\infty}\f{1}{n!}\sum_{T\in\mathcal{T}_n}\left(\prod_{l\in T}\int_0^1dh_l\right)\left(\prod_{l\in T}\f{\partial}{\partial h_l}\right)\int d\nu^{\boldsymbol{h^T}}_n(\sigma_1,\dots,\sigma_n)\prod_{v=1}^nV_{\lambda}(\sigma_v).
\ee
It is this tree $T$ over loop vertices $V_{\lambda}(\sigma_{v})$ which we coin a `cactus'.

Since the dependence of the covariance of $d\nu^{\boldsymbol{h^T}}_n(\sigma_1,\dots,\sigma_n)$
in the $h$ variables is linear, applying the derivative $\f{\partial}{\partial h_l}$ in \ref{toytrees} is easy using (\ref{expquadr}):
it amounts to an additional insertions of $\f{\partial^2}{\partial\sigma_{s(l)}\partial\sigma_{t(l)}}$ in the integral, where $s(l)$ and $t(l)$
are respectively the starting and ending vertices of the line $l$. Hence
\be\label{derivatexp}
\left(\prod_{l\in T}\f{\partial}{\partial h_l}\right)\int d\nu^{\boldsymbol{h^T}}_n(\sigma_1,\dots,\sigma_n)\prod_{v=1}^nV_{\lambda}(\sigma_v)=\int d\nu^{\boldsymbol{h^T}}_n(\sigma_1,\dots,\sigma_n)\left(\prod_{l\in T}\f{\partial^2}{\partial\sigma_{s(l)}\partial\sigma_{t(l)}}\right)\prod_{v=1}^nV_{\lambda}(\sigma_v).
\ee

Consider the loop vertex $V_{\lambda}(\sigma_v)$, with coordination $k_v$ in the tree $T$. Thanks to the resolvent bound, we have

\be\label{bound}
\vert\f{\partial^{k_v}}{\partial\sigma_v^{k_v}}V_{\lambda}(\sigma_v)\vert=(k_v-1)!\vert\lambda\vert^{\f{k_v}{2}}\vert1+i\sqrt{\lambda}\sigma_v\vert^{-k_v}\leq2^{\f{k_v}{2}}(k_v-1)!\vert\lambda\vert^{\f{k_v}{2}},
\ee
and thus, since there are $n-1$ lines in a tree over $n$ vertices,
\be
\vert\left(\prod_{l\in T}\f{\partial^2}{\partial\sigma_{s(l)}\partial\sigma_{t(l)}}\right)\prod_{v=1}^nV_{\lambda}(\sigma_v)\vert\leq2^{n-1}\vert\lambda\vert^{n-1}\prod_{v=1}^n(k_v-1)!
\ee
This bound goes through the normalized integrals over the $\sigma$'s and the $h$'s, and using Cayley's formula for the number of trees over $n$ labeled vertices (\ref{cayley}), we find that the summand in (\ref{toytrees}) is bounded by $2^{n-1}\vert\lambda\vert^{n-1}$. This shows that the cactus expansion (\ref{toytrees}) of $F$ converges uniformly in a half-disk $D_{R}=\{\lambda\in\mathbb{C},\Re\lambda\geq0,\vert\lambda\vert\leq R\}$, with $R<\f{1}{2}$.

%\footnote{There is a little subtlety with the first term in $F$, for $n=1$, which is a single
%loop vertex with value $\log (1 + i \sqrt{\lambda} \sigma )$.} To transform it also into an 
%expression with denominators, one should perform a single Taylor expansion step
%and integrate the $\sigma$ field:
%\bea \int d \nu (\sigma)  \log [ 1 + i \sqrt{\lambda} \sigma ] &=& \int d \nu (\sigma)
%\int_0^1 dt  \frac {i \sqrt{\lambda} \sigma }
%{ 1 + i \sqrt{\lambda} t \sigma } \nonumber= \int d \nu (\sigma)
%\int_0^1 dt  \frac {  \lambda t} { [1 + i \sqrt{\lambda} t \sigma ]^2 

%This application of the resolvent bound, however, relies on the fact that $T$ has at least one link, and we must also consider the trivial tree with a single loop vertex $\text{Log}(1+i\sqrt{\lambda}\sigma)$ and no link. To bound it as well, we perform a single Taylor expansion step, apply the resolvent bound and integrate the $\sigma$ `field' 

%\be
%\vert\int d\nu(\sigma)\ \text{Log}(1+i\sqrt{\lambda}\sigma)\vert=\vert\int d\nu(\sigma)\int_0^1 dt\ \f{\lambda\sigma^2t}{\vert1+i\sqrt{\lambda}\sigma t\vert^2}\vert\leq\f{\sqrt{2}}{2}\vert\lambda\vert.
%\ee

Let us emphasize here that this result does not contradict the field-theoretic wisdom, which goes back to Dyson \cite{PhysRev.85.631}, that a typical QFT perturbative series is bound to diverge: the lemma that convergence at one point implies analyticity in a disk around the origin, on which Dyson's argument relies, only applies to {\it power} series -- and the cactus expansion (\ref{sum}) is {\it not} a power series. 

Of course, one might wonder what the convergence of the cactus expansion teaches us about the {\it perturbative} expansion of 
$F(\lambda)$

\be\label{feynman}
F(\lambda)\simeq\sum_{p\geq0}(-\lambda)^pa_p
\ee
which, {\it in fine}, is the one dealt with in standard QFT. The answer is simple: the cactus expansion proves Borel summability of the perturbative series, and gives an explicit expression for its Borel sum (Appendix C). Indeed, since $V_{\lambda}$ is analytic in the cut plane, and $F$ is the sum of a power series in $V_{\lambda}$ converging uniformly in the half-disk $D_{R}$, we now know that $F$ is analytic in a Nevanlinna-Sokal disk $C_{R/2}$. %What is more, expanding the product of resolvents $\prod_v(1+i\sqrt{\lambda}\sigma_v)^{-k_v}$ in power of $\lambda$ (keeping only even powers, since the Gaussian integration over the intermediate fields $\sigma_v$ cancels any odd monomial in the $\sigma_v$'s), we recover the Feynman expansion (\ref{feynman}) order by order.
Moreover, the order-$r$ Taylor-Lagrange remainder $T_rF(\lambda):= F(\lambda)- \sum_{p=0}^{r-1} 
\frac{\lambda^p}{p!} F^{(p)} (0)$ can easily be shown to satisfy the Nevanlinna-Sokal criterion (\ref{bor2}). Indeed, consider a cactus 
amplitude $\cC_T$ with $n$ loop vertices:
\be
\cC_T:=\left(\prod_{l\in T}\int_0^1dh_l\right)\left(\prod_{l\in T}\f{\partial}{\partial h_l}\right)\int d\nu^{\boldsymbol{h^T}}_n(\sigma_1,\dots,\sigma_n)\prod_{v=1}^nV_{\lambda}(\sigma_v).
\ee

By (\ref{derivatexp})-(\ref{bound}), each such amplitude is made of an explicit factor $\lambda^{n-1}$ times an integral over $d\nu^{\boldsymbol{h^T}}_n(\sigma_1,\dots,\sigma_n)$ of
a product of $2n-2$ resolvents
$R_{l_{v}} (\lambda, \sigma_{v})  :=(1+i\sqrt{\lambda}\sigma_{v})^{-1}$, where $l_{v}$ denote a half-line hooked to a vertex $v$. Hence for $r \le n-1$, $T_r  \cC_T =  \cC_T $, and for $r\ge n$
\be  T_r \cC_T   =  \lambda^{n-1} T_{r-n+1}
 \left(\prod_{l\in T}\int_0^1dh_l\right)\int d\nu^{\boldsymbol{h^T}}_n(\sigma_1,\dots,\sigma_n) \prod_{l_{v}=1}^{2n-2} R_{l_{v}} (\lambda, \sigma_{v}) .
\ee
But by (\ref{expquadr}) and since $F=  \prod_{l_{v}=1}^{2n-2} R_{l_{v}}$ is solely a function of $\sqrt{\lambda} \sigma $
we have
\be
\int d\nu^{\boldsymbol{h^T}}_n(\sigma_1,\dots,\sigma_n)  \prod_{l_{v}=1}^{2n-2} R_{l_{v}} (\lambda, \sigma_{v}) = 
e^{\frac{1}{2}\lambda \frac{\partial}{\partial \sigma}  C^{\boldsymbol{h}^T} \frac{\partial}{\partial \sigma }  }  
 \prod_{l_{v}=1}^{2n-2}  R_{l_{v}} (1, \sigma)  \; \vert_{\sigma_{v} =0}.
\ee 
Hence the Taylor-Lagrange formula applies to the exponential:
\be  T_k e^{\lambda H}  = \int_0^1 dt\ \frac{(1-t)^{k-1}}{(k-1)!}   \lambda^{k}  H^{k} e^{\lambda t H} 
\ee
and the operator $H^k$ creates exactly $k$ additional insertions of 
lines $\f{\partial^2}{\partial\sigma_{s(l')}\partial\sigma_{t(l')}}$. The combinatorics of $2k$ derivations 
on a product of $2n-2$ resolvents costs a factor $\frac{(2r-1))!}{(2n-1)!}$
times a new Gaussian integral with $2n-2+2k$ resolvents of the same type $R_{l_{v}'}$. The $\lambda$
factor can be then transferred back to a $\sqrt\lambda$ factor in the resolvents:
\be  e^{\frac{1}{2}\lambda t \frac{\partial}{\partial \sigma}  C^{\boldsymbol{h}^T} \frac{\partial}{\partial \sigma }  }   \prod_{l_{v}'=1}^{2n-2+2k} 
R_{l_{v}'} (1, \sigma_{v}) \; \vert_{\sigma_{v} =0} =  e^{\frac{1}{2}t \frac{\partial}{\partial \sigma}  C^{\boldsymbol{h}^T} \frac{\partial}{\partial \sigma }  }   
\prod_{l_{v}'=1}^{2n-2+2k}  R_{l_{v}'} (\lambda, \sigma_{v}) \; \vert_{\sigma_{v} =0}  .
\ee
Hence since $k = r-n+1$ the convergent series $\sum_{n=1}^{\infty}\f{1}{n!}\sum_{T\in\mathcal{T}_n} T_r \cC_T $ can be bounded exactly as before,
except for two facts: each term contains a factor $\lambda^{r}$ and we have also to add to the bounds a factor 
$\int_0^1 dt \frac{(1-t)^{r-n}}{(r-n)!} \frac{(2r-1))!}{(2n-1)!}=\frac{(2r-1))!}{(r-n+1)!(2n-1)!}$.
This last factor is maximal for the trivial tree with $n=1$,  and certainly bounded by $2^r\, r!$, from 
which it follows that for some constant $K$
\be
\vert T_r  F(\lambda)\vert\leq K^r  r!\vert\lambda\vert^r.
\ee

To summarize, the cactus expansion allows not only to trade the {\it asymptotic} perturbative series (\ref{feynman}) for the {\it convergent} expression (\ref{toytrees}), but also to check the Sokal-Nevanlinna criteria, proving Borel summability of the former. Of course, in this toy example, Borel summability of $Z$ is obvious and Borel summability of $F=\log Z$ could be shown by more elementary methods; the power of the cactus expansion becomes manifest when it comes to the constructive analysis of the $\phi^4_4$ field theory \cite{Magnen:2007uy}, and of the matrix $\phi^4$ model \cite{Rivasseau:2007fr}. Our aim in this paper is to extend its scope to tensor models such as Boulatov's GFT.

%%%%%%%%%%%%%%%%%%%%%%%%%
\section{Construction of the BFL model}
%%%%%%%%%%%%%%%%%%%%%%%%%

\subsection{Intermediate field representation}

\begin{figure}[h]
\begin{center}
\scalebox{0.7} % Change this value to rescale the drawing.
{
\begin{pspicture}(0,-2.056)(9.024,2.016)
\psbezier[linewidth=0.024](5.4166675,1.1479173)(6.212,0.604)(6.212,0.0039999997)(5.432209,-0.54906785)
\psbezier[linewidth=0.024](6.412,-0.196)(6.212,-0.196)(5.812,-0.596)(5.432209,-0.9490679)
\psbezier[linewidth=0.024](6.412,-0.596)(6.012,-0.596)(5.612,-1.196)(5.432209,-1.3490678)
\psbezier[linewidth=0.024](6.412,1.004)(6.012,1.204)(5.812,1.604)(5.412,2.004)
\psbezier[linewidth=0.024](6.412,0.604)(6.212,0.604)(5.612,1.404)(5.412,1.604)
\psline[linewidth=0.04cm,linestyle=dashed,dash=0.16cm 0.16cm](6.412,1.004)(8.012,1.004)
\psline[linewidth=0.04cm,linestyle=dashed,dash=0.16cm 0.16cm](6.412,0.604)(8.012,0.604)
\psline[linewidth=0.04cm,linestyle=dashed,dash=0.16cm 0.16cm](6.412,-0.196)(8.012,-0.196)
\psline[linewidth=0.04cm,linestyle=dashed,dash=0.16cm 0.16cm](6.412,-0.596)(8.012,-0.596)
\psbezier[linewidth=0.024](9.007333,1.1479173)(8.212,0.604)(8.212,0.0039999997)(8.991791,-0.54906785)
\psbezier[linewidth=0.024](8.012,-0.196)(8.212,-0.196)(8.612,-0.596)(8.991791,-0.9490679)
\psbezier[linewidth=0.024](8.012,-0.596)(8.412,-0.596)(8.812,-1.196)(8.991791,-1.3490678)
\psbezier[linewidth=0.024](8.012,1.004)(8.412,1.204)(8.612,1.604)(9.012,2.004)
\psbezier[linewidth=0.024](8.012,0.604)(8.212,0.604)(8.812,1.404)(9.012,1.604)
\psbezier[linewidth=0.024](0.016667435,1.1479173)(0.812,0.604)(0.812,0.0039999997)(0.03220895,-0.54906785)
\psbezier[linewidth=0.024](1.012,-0.196)(0.612,-0.396)(0.212,-0.796)(0.03220895,-0.9490679)
\psbezier[linewidth=0.024](1.012,-0.596)(0.612,-0.596)(0.212,-1.196)(0.03220895,-1.3490678)
\psbezier[linewidth=0.024](1.012,1.004)(0.612,1.204)(0.412,1.604)(0.012,2.004)
\psbezier[linewidth=0.024](1.012,0.604)(0.612,0.804)(0.212,1.404)(0.012,1.604)
\psline[linewidth=0.04cm,linestyle=dashed,dash=0.16cm 0.16cm](1.012,1.004)(2.612,1.004)
\psline[linewidth=0.04cm,linestyle=dashed,dash=0.16cm 0.16cm](1.012,0.604)(2.612,-0.196)
\psline[linewidth=0.04cm,linestyle=dashed,dash=0.16cm 0.16cm](1.012,-0.196)(2.612,0.604)
\psline[linewidth=0.04cm,linestyle=dashed,dash=0.16cm 0.16cm](1.012,-0.596)(2.612,-0.596)
\psbezier[linewidth=0.024](3.6073325,1.1479173)(2.812,0.604)(2.812,0.0039999997)(3.5917912,-0.54906785)
\psbezier[linewidth=0.024](2.612,-0.196)(3.012,-0.396)(3.412,-0.796)(3.5917912,-0.9490679)
\psbezier[linewidth=0.024](2.612,-0.596)(3.012,-0.596)(3.412,-1.196)(3.5917912,-1.3490678)
\psbezier[linewidth=0.024](2.612,1.004)(3.012,1.204)(3.212,1.604)(3.612,2.004)
\psbezier[linewidth=0.024](2.612,0.604)(3.012,0.804)(3.412,1.404)(3.612,1.604)
\usefont{T1}{ptm}{m}{n}
\rput(7.282,-1.891){Pillow}
\usefont{T1}{ptm}{m}{n}
\rput(1.892,-1.891){Tetrahedron}
\end{pspicture} 
}
\end{center}
\caption{Slicing the BFL vertices with an intermediate field $\sigma$ over $\su^4$: the dashed lines are combined in the covariance $C_{\sigma}$.}
\label{split}
\end{figure}
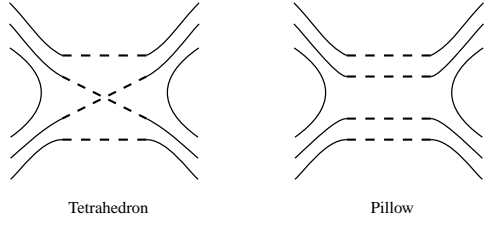

Let us now construct the cactus expansion of the BFL model. Following the recipe explained above, we first introduce a ultralocal intermediate field $\sigma$ on $\su^4$ slicing the BFL $\phi^4$ vertices into two $\phi^2\sigma$ vertices as in Fig. \ref{split}:
\be
e^{-\lambda I_{\delta}[\phi]/8}=\int d\nu^{(\La)}_{\delta}[\sigma]\ e^{-\f{i}{2}\sqrt{\lambda}\la S\phi\vert\sigma\ra_4}.
\ee
Note that, in this intermediate field picture, the tetrahedral and pillow interactions are encapsulated in the ultralocal Gaussian measure $d\nu^{(\La)}_{\delta}$ through its covariance $C_{\sigma}:=(1+\delta\mathcal{T})$. 
% Generated with LaTeXDraw 2.0.2
% Tue Jun 23 12:49:38 CEST 2009
% \usepackage[usenames,dvipsnames]{pstricks}
% \usepackage{epsfig}
% \usepackage{pst-grad} % For gradients
% \usepackage{pst-plot} % For axes

Introducing the operator $\Sigma$ coupling $\phi$ to $\sigma$
\be
\Sigma\phi(g_1,g_2,g_3):=\int dg_4dg_5\ \sigma(g_1,g_2,g_4,g_5)\phi(g_3,g_4,g_5)
\ee
in such a way that
\be
\la S\phi\vert\sigma\ra_4=\la\phi\vert\Sigma\phi\ra_3,
\ee
where $\la\cdot\vert\cdot\ra_{3}$ is the standard inner product in $L^2(\su^3)$, we obtain after integration over the original field $\phi$
\be
\mathcal{Z}_{\text{BFL}}^{(\La)}(\lambda)=\int d\nu^{(\La)}_{\delta}[\sigma]\ e^{V_{\lambda}[\sigma]},
\ee
where the `loop vertex' is given by 
\be
V_{\lambda}[\sigma]:=-\f{1}{2} \textrm{Tr}\ \textrm{Log}(1+i\sqrt{\lambda}C_{\phi}\Sigma C_{\phi}).
\ee

One easily checks that $\widetilde{\Sigma}:=C_{\phi}\Sigma C_{\phi}$ is a Hermitian operator and therefore that the resolvent bound applies to the derivatives of this `loop vertex' just like in the toy example.

\subsection{Cactus expansion}

Following the same steps as in sec. IV yields the cactus expansion of the BFL free energy:

\begin{eqnarray}\label{cactusBFL}
\mathcal{F}^{(\La)}_{\text{BFL}}(\lambda)=\sum_{n=1}^{\infty}\f{1}{n!}&&\sum_{T\in\mathcal{T}_n}\left(\prod_{l\in T}\int_0^1dh_l\right)\int d\nu^{\boldsymbol{h^T}}_n(\sigma_1,\dots,\sigma_n)\nonumber\\
&&\left(\prod_{l\in T}\int d^4g^{s(l)}d^4g^{t(l)}\ C_{\sigma}(g^{s(l)};g^{t(l)})\f{\delta^2}{\delta\sigma_{s(l)}(g^{s(l)})\delta\sigma_{t(l)}(g^{t(l)})}\right)\prod_{v=1}^nV_{\lambda}(\sigma_v).
\end{eqnarray}

At this stage, the only difference with the 0-dimensional case is the insertion of a covariance $C_{\sigma}(g_{s(l)};g_{t(l)})$ on each line $l$ of the tree, and the integration with respect to the corresponding $4$-uples of group elements $g^{l_v}:=(g_i^{l_v})_{i=1}^4$, attached to the half-lines $l_{v}$.

Computing the effect of the $k_v$ derivatives on the loop vertex $V_{\lambda}[\sigma_v]$, labeled by the half-lines $l_v$ connecting it to the tree, we obtain
\be
\left(\prod_{l_v=1}^{k_v}\f{\delta}{\delta\sigma_{v}(g^{l_v})}\right)V_{\lambda}(\sigma_v)=\f{(i\sqrt{\lambda})^{k_v}}{2}\textrm{Tr}\left(\prod_{l_v=1}^{k_v}(1+i\sqrt{\lambda}\widetilde{\Sigma}(\sigma_v))^{-1}\f{\delta\widetilde{\Sigma}}{\delta\sigma_v(g^{l_v})}\right),
\ee
and the cactus amplitude is given by the following product of traces connected by ultralocal covariances:
\bea\label{bigprod}
\cC_T(\lambda):=&&\left(\prod_{l\in T}\int_0^1dh_l\right)\int d\nu^{\boldsymbol{h^T}}_n(\sigma_1,\dots,\sigma_n)\nonumber\\ &&\prod_{l\in T}\int d^4g^{s(l)}d^4g^{t(l)}\ C_{\sigma}(g^{s(l)};g^{t(l)})\prod_{v\in T}\textrm{Tr}\left(\prod_{l_v=1}^{k_v}(1+i\sqrt{\lambda}\widetilde{\Sigma}(\sigma_v))^{-1}\f{\delta\widetilde{\Sigma}}{\delta\sigma_v(g^{l_v})}\right).
\eea
%where $A_{l_v}(g^{l_v}):=(1+i\sqrt{\lambda}\widetilde{\Sigma}(\sigma_v))^{-1}\f{\delta\widetilde{\Sigma}}{\delta\sigma_v(g^{l_v})}$. 

\subsection{Cauchy-Schwarz inequalities}

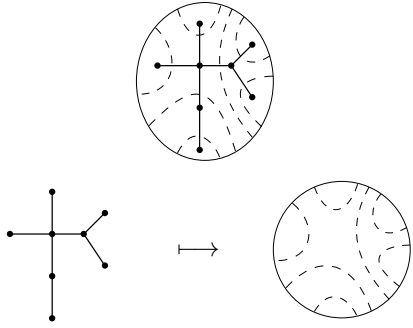
\begin{figure}[h]
\begin{center}
\scalebox{0.7} % Change this value to rescale the drawing.
{
\begin{pspicture}(0,-3.036)(7.681,3.043)
\psdots[dotsize=0.12](3.672,2.636)
\psdots[dotsize=0.12](3.672,1.836)
\psdots[dotsize=0.12](2.872,1.836)
\psdots[dotsize=0.12](4.272,1.836)
\psdots[dotsize=0.12](4.672,2.236)
\psdots[dotsize=0.12](4.672,1.236)
\psdots[dotsize=0.12](3.672,1.036)
\psdots[dotsize=0.12](3.672,0.236)
\psline[linewidth=0.024cm](3.672,2.636)(3.672,1.836)
\psline[linewidth=0.024cm](3.672,1.836)(2.872,1.836)
\psline[linewidth=0.024cm](3.672,1.836)(3.672,1.036)
\psline[linewidth=0.024cm](3.672,1.836)(4.272,1.836)
\psline[linewidth=0.024cm](4.272,1.836)(4.672,2.236)
\psline[linewidth=0.024cm](4.272,1.836)(4.672,1.236)
\psline[linewidth=0.024cm](3.672,1.036)(3.672,0.236)
\psellipse[linewidth=0.0139999995,dimen=middle](3.772,1.536)(1.3,1.5)
\psbezier[linewidth=0.018,linestyle=dashed,dash=0.16cm 0.16cm](4.512,2.756)(4.192,2.3256774)(4.397714,1.616)(5.012,2.0263226)
\psbezier[linewidth=0.018,linestyle=dashed,dash=0.16cm 0.16cm](4.052,0.096)(3.932,0.416)(3.552,0.776)(3.252,0.176)
\psbezier[linewidth=0.018,linestyle=dashed,dash=0.16cm 0.16cm](2.8334925,2.556)(3.332,2.096)(3.232,1.316)(2.572,1.296)
\psbezier[linewidth=0.018,linestyle=dashed,dash=0.16cm 0.16cm](3.252,2.916)(3.457,2.236)(3.872,2.236)(4.072,2.976)
\psbezier[linewidth=0.018,linestyle=dashed,dash=0.16cm 0.16cm](2.712,0.696)(3.512,1.516)(3.932,1.616)(4.352,0.216)
\psbezier[linewidth=0.018,linestyle=dashed,dash=0.16cm 0.16cm](4.292,2.916)(3.932,2.156)(3.912,1.376)(4.672,0.476)
\psbezier[linewidth=0.018,linestyle=dashed,dash=0.16cm 0.16cm](5.072,1.636)(4.672,1.636)(4.012,1.456)(4.852,0.676)
\psellipse[linewidth=0.0139999995,dimen=middle](6.372,-1.661001)(1.3,1.2970009)
\psbezier[linewidth=0.018,linestyle=dashed,dash=0.16cm 0.16cm](7.112,-0.6061068)(6.792,-0.9781927)(6.9977145,-1.5918275)(7.612,-1.237035)
\psbezier[linewidth=0.018,linestyle=dashed,dash=0.16cm 0.16cm](6.652,-2.9061217)(6.532,-2.6294281)(6.152,-2.3181481)(5.852,-2.8369484)
\psbezier[linewidth=0.018,linestyle=dashed,dash=0.16cm 0.16cm](5.4334927,-0.7790403)(5.932,-1.1767873)(5.832,-1.8512278)(5.172,-1.8685211)
\psbezier[linewidth=0.018,linestyle=dashed,dash=0.16cm 0.16cm](5.852,-0.4677601)(6.057,-1.0557338)(6.472,-1.0557338)(6.672,-0.41588002)
\psbezier[linewidth=0.018,linestyle=dashed,dash=0.16cm 0.16cm](5.312,-2.3873215)(5.992,-1.6437075)(6.632,-1.9420543)(6.932,-2.824)
\psbezier[linewidth=0.018,linestyle=dashed,dash=0.16cm 0.16cm](6.892,-0.4677601)(6.532,-1.1249073)(6.512,-1.7712277)(7.272,-2.604)
\psbezier[linewidth=0.018,linestyle=dashed,dash=0.16cm 0.16cm](7.672,-1.5745342)(7.272,-1.5745342)(6.672,-1.7474676)(7.472,-2.364)
\psdots[dotsize=0.12](0.872,-0.564)
\psdots[dotsize=0.12](0.872,-1.364)
\psdots[dotsize=0.12](0.072,-1.364)
\psdots[dotsize=0.12](1.472,-1.364)
\psdots[dotsize=0.12](1.872,-0.964)
\psdots[dotsize=0.12](1.872,-1.964)
\psdots[dotsize=0.12](0.872,-2.164)
\psdots[dotsize=0.12](0.872,-2.964)
\psline[linewidth=0.024cm](0.872,-0.564)(0.872,-1.364)
\psline[linewidth=0.024cm](0.872,-1.364)(0.072,-1.364)
\psline[linewidth=0.024cm](0.872,-1.364)(0.872,-2.164)
\psline[linewidth=0.024cm](0.872,-1.364)(1.472,-1.364)
\psline[linewidth=0.024cm](1.472,-1.364)(1.872,-0.964)
\psline[linewidth=0.024cm](1.472,-1.364)(1.872,-1.964)
\psline[linewidth=0.024cm](0.872,-2.164)(0.872,-2.964)
\usefont{T1}{ptm}{m}{n}
\rput(3.652,-1.679){\Large $\longmapsto$}
\end{pspicture} 
}
\end{center}
\caption{The planar representation of a tree.}
\label{disk}
\end{figure}

To get some insight into this cactus amplitude, it is handy to introduce a `dual' representation of a tree $T$, as a planar partition of the disk. The boundary of the disk is obtained by turning around $T$, while the dotted lines partitioning it cross the boundary twice and each line of $T$ exactly once, without crossing each other, see Fig. \ref{disk}.

% Generated with LaTeXDraw 2.0.2
% Sat Jun 13 15:15:26 CEST 2009
% \usepackage[usenames,dvipsnames]{pstricks}
% \usepackage{epsfig}
% \usepackage{pst-grad} % For gradients
% \usepackage{pst-plot} % For axes

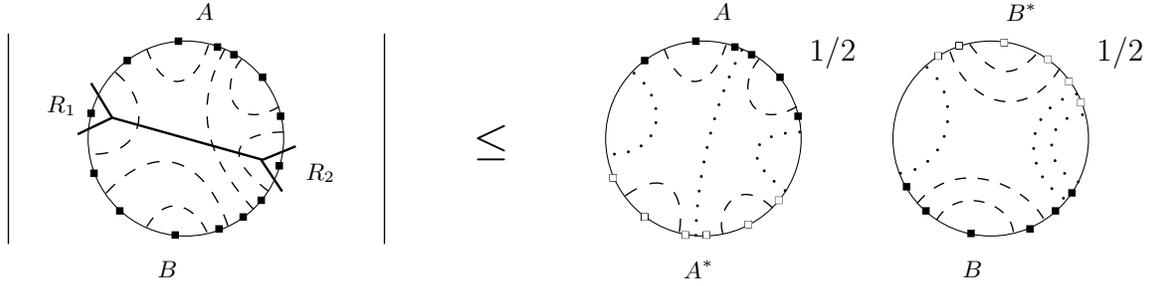
\begin{figure}[h]
\begin{center}
\scalebox{1} % Change this value to rescale the drawing.
{
\begin{pspicture}(0,-1.9)(15.32,1.9)
\psellipse[linewidth=0.0139999995,dimen=middle](2.36,0.022999091)(1.3,1.2970009)
\psbezier[linewidth=0.018,linestyle=dashed,dash=0.16cm 0.16cm](3.1,1.0778931)(2.78,0.7058073)(2.9857142,0.09217247)(3.6,0.44696498)
\psbezier[linewidth=0.018,linestyle=dashed,dash=0.16cm 0.16cm](2.64,-1.2221218)(2.52,-0.94542825)(2.14,-0.63414806)(1.84,-1.1529484)
\psbezier[linewidth=0.018,linestyle=dashed,dash=0.16cm 0.16cm](1.4214926,0.90495974)(1.92,0.50721276)(1.82,-0.16722772)(1.16,-0.18452105)
\psbezier[linewidth=0.018,linestyle=dashed,dash=0.16cm 0.16cm](1.84,1.2162399)(2.045,0.62826616)(2.46,0.62826616)(2.66,1.2681199)
\psbezier[linewidth=0.018,linestyle=dashed,dash=0.16cm 0.16cm](1.3,-0.7033214)(1.98,0.040292434)(2.62,-0.25805432)(2.92,-1.14)
\psbezier[linewidth=0.018,linestyle=dashed,dash=0.16cm 0.16cm](2.88,1.2162399)(2.52,0.5590928)(2.5,-0.08722771)(3.26,-0.92)
\psbezier[linewidth=0.018,linestyle=dashed,dash=0.16cm 0.16cm](3.66,0.109465815)(3.26,0.109465815)(2.66,-0.06346764)(3.46,-0.68)
\psline[linewidth=0.032cm](1.38,0.3)(3.38,-0.26)
\psline[linewidth=0.032cm](1.1,0.76)(1.38,0.3)
\psline[linewidth=0.032cm](1.38,0.3)(0.92,0.08)
\psline[linewidth=0.032cm](3.38,-0.26)(3.82,-0.08)
\psline[linewidth=0.032cm](3.36,-0.26)(3.64,-0.68)
\psdots[dotsize=0.11,dotstyle=square*](2.26,1.32)
\psdots[dotsize=0.11,dotstyle=square*](2.78,1.24)
\psdots[dotsize=0.11,dotstyle=square*](3.0,1.14)
\psdots[dotsize=0.11,dotstyle=square*](3.38,0.84)
\psdots[dotsize=0.11,dotstyle=square*](3.62,0.32)
\psdots[dotsize=0.11,dotstyle=square*](3.6,-0.34)
\psdots[dotsize=0.11,dotstyle=square*](3.36,-0.8)
\psdots[dotsize=0.11,dotstyle=square*](3.12,-1.02)
\psdots[dotsize=0.11,dotstyle=square*](2.8,-1.18)
\psdots[dotsize=0.11,dotstyle=square*](2.22,-1.26)
\psdots[dotsize=0.11,dotstyle=square*](1.48,-0.94)
\psdots[dotsize=0.11,dotstyle=square*](1.14,-0.44)
\psdots[dotsize=0.11,dotstyle=square*](1.58,1.06)
\psdots[dotsize=0.11,dotstyle=square*](1.1,0.36)
\usefont{T1}{ptm}{m}{n}
\rput(2.12,-1.715){$B$}
\usefont{T1}{ptm}{m}{n}
\rput(0.71,0.465){\small $R_1$}
\usefont{T1}{ptm}{m}{n}
\rput(2.61,1.705){$A$}
\usefont{T1}{ptm}{m}{n}
\rput(4.15,-0.435){\small $R_2$}
\psline[linewidth=0.018cm](0.0,1.42)(0.0,-1.38)
\psline[linewidth=0.018cm](5.0,1.42)(5.0,-1.38)
\usefont{T1}{ptm}{m}{n}
\rput(6.42,-0.025){\LARGE $\leq$}
\psellipse[linewidth=0.0139999995,dimen=middle](9.24,0.022999091)(1.3,1.2970009)
\psbezier[linewidth=0.018,linestyle=dashed,dash=0.16cm 0.16cm](9.98,1.0778931)(9.66,0.7058073)(9.865714,0.09217247)(10.48,0.44696498)
\psbezier[linewidth=0.04,linestyle=dotted,dotsep=0.16cm](8.301493,0.90495974)(8.8,0.50721276)(8.7,-0.16722772)(8.04,-0.18452105)
\psbezier[linewidth=0.018,linestyle=dashed,dash=0.16cm 0.16cm](8.72,1.2162399)(8.925,0.62826616)(9.34,0.62826616)(9.54,1.2681199)
\psbezier[linewidth=0.04,linestyle=dotted,dotsep=0.16cm](9.76,1.2162399)(9.4,0.5590928)(9.14,-0.5)(9.14,-1.28)
\psbezier[linewidth=0.04,linestyle=dotted,dotsep=0.16cm](10.54,0.109465815)(10.14,0.109465815)(9.54,-0.06346764)(10.34,-0.68)
\psdots[dotsize=0.11,dotstyle=square*](9.14,1.32)
\psdots[dotsize=0.11,dotstyle=square*](9.66,1.24)
\psdots[dotsize=0.11,dotstyle=square*](9.88,1.14)
\psdots[dotsize=0.11,dotstyle=square*](10.26,0.84)
\psdots[dotsize=0.11,dotstyle=square*](10.5,0.32)
\psdots[dotsize=0.11,dotstyle=square*](8.46,1.06)
\usefont{T1}{ptm}{m}{n}
\rput(9.49,1.705){$A$}
\usefont{T1}{ptm}{m}{n}
\rput(9.18,-1.715){$A^*$}
\psbezier[linewidth=0.018,linestyle=dashed,dash=0.16cm 0.16cm](10.16,-0.9)(9.68,-0.44)(9.4,-0.98)(9.5,-1.24)
\psbezier[linewidth=0.018,linestyle=dashed,dash=0.16cm 0.16cm](8.2,-0.74)(8.78,-0.32)(9.0,-0.74)(8.92,-1.22)
\psdots[dotsize=0.11,fillstyle=solid,dotstyle=square](8.04,-0.5)
\psdots[dotsize=0.11,fillstyle=solid,dotstyle=square](8.46,-1.02)
\psdots[dotsize=0.11,fillstyle=solid,dotstyle=square](9.0,-1.26)
\psdots[dotsize=0.11,fillstyle=solid,dotstyle=square](9.28,-1.26)
\psdots[dotsize=0.11,fillstyle=solid,dotstyle=square](9.84,-1.12)
\psdots[dotsize=0.11,fillstyle=solid,dotstyle=square](10.24,-0.8)
\usefont{T1}{ptm}{m}{n}
\rput(10.96,1.17){\large $1/2$}
\psellipse[linewidth=0.0139999995,dimen=middle](13.06,0.022999091)(1.3,1.2970009)
\usefont{T1}{ptm}{m}{n}
\rput(13.47,1.705){$B^*$}
\usefont{T1}{ptm}{m}{n}
\rput(12.82,-1.715){$B$}
\usefont{T1}{ptm}{m}{n}
\rput(14.78,1.17){\large $1/2$}
\psbezier[linewidth=0.018,linestyle=dashed,dash=0.16cm 0.16cm](12.34,-1.06)(12.62,-0.6)(13.32,-0.82)(13.36,-1.22)
\psbezier[linewidth=0.018,linestyle=dashed,dash=0.16cm 0.16cm](12.08,-0.8423529)(12.6,-0.2)(13.5,-0.58)(13.76,-1.04)
\psbezier[linewidth=0.04,linestyle=dotted,dotsep=0.16cm](11.84,-0.44)(12.48,-0.4)(12.74,0.54)(12.2,0.98)
\psbezier[linewidth=0.04,linestyle=dotted,dotsep=0.16cm](14.22,0.62)(13.52,0.46)(13.42,-0.38)(14.04,-0.78)
\psbezier[linewidth=0.04,linestyle=dotted,dotsep=0.16cm](14.3,0.34)(13.8,0.36)(13.68,-0.34)(14.24,-0.56)
\psbezier[linewidth=0.018,linestyle=dashed,dash=0.16cm 0.16cm](12.52,1.18)(12.72,0.5)(13.52,0.2)(13.98,0.96)
\psbezier[linewidth=0.018,linestyle=dashed,dash=0.16cm 0.16cm](12.78,1.28)(13.1,0.82)(13.36,0.82)(13.66,1.14)
\psdots[dotsize=0.11,fillstyle=solid,dotstyle=square](12.36,1.12)
\psdots[dotsize=0.11,fillstyle=solid,dotstyle=square](12.64,1.26)
\psdots[dotsize=0.11,fillstyle=solid,dotstyle=square](13.24,1.3)
\psdots[dotsize=0.11,fillstyle=solid,dotstyle=square](13.82,1.08)
\psdots[dotsize=0.11,fillstyle=solid,dotstyle=square](14.1,0.78)
\psdots[dotsize=0.11,fillstyle=solid,dotstyle=square](14.26,0.5)
\psdots[dotsize=0.11,dotstyle=square*](11.94,-0.62)
\psdots[dotsize=0.11,dotstyle=square*](12.2,-0.94)
\psdots[dotsize=0.11,dotstyle=square*](12.8,-1.24)
\psdots[dotsize=0.11,dotstyle=square*](13.58,-1.18)
\psdots[dotsize=0.11,dotstyle=square*](13.92,-0.94)
\psdots[dotsize=0.11,dotstyle=square*](14.14,-0.7)
\end{pspicture} 
}
\end{center}
\caption{Splitting the disk in two parts to apply the Cauchy-Schwarz inequality. On the LHS, black squares are resolvents, dashed lines are covariances $C_{\sigma}$, and the thick line expresses the amplitude (\ref{bigprod}) as the inner product between the upper and lower parts $A$ and $B$. On the right, white squares are Hermitian conjugates of resolvents and dotted lines are covariances $C_{\sigma}'$.}
\label{schwarz1}
\end{figure}

In such a picture, the resolvents $(1+i\sqrt{\lambda}\widetilde{\Sigma})^{-1}$ are attached to the arcs on the boundary of the disk, while the covariances $C_{\sigma}$ are attached to the dotted lines. To bound (\ref{bigprod}), we can apply the Cauchy-Schwarz inequality along a line splitting the disk in two parts with the same number of consecutive resolvents. Indeed, the number of half-lines of a tree being even, it is always possible to pick two arcs with resolvents $R_1$ and $R_2$, and express (\ref{bigprod}) as the inner product $\la A\vert R_1\otimes R_2\vert B\ra$, where $A$ and $B$ contain the same number of arcs, and thus of resolvents (see Fig. \ref{schwarz1}). By the Cauchy-Schwarz inequality, we have

\be
\vert\la A\vert R_1\otimes R_2\vert B\ra\vert\leq\Vert R_1\Vert\vert R_2\Vert\sqrt{\la A\vert A\ra}\sqrt{\la B\vert B\ra}\leq 2\sqrt{\la A\vert A\ra}\sqrt{\la B\vert B\ra}.
\ee

\begin{figure}[h]
\begin{center}
\scalebox{0.7} % Change this value to rescale the drawing.
{
\begin{pspicture}(0,-1.6885338)(21.752,1.6885339)
\psbezier[linewidth=0.024](5.5446677,0.82045114)(6.34,0.27653393)(6.34,-0.32346618)(5.560209,-0.876534)
\psbezier[linewidth=0.024](6.54,-0.52346605)(6.34,-0.52346605)(5.94,-0.9234661)(5.560209,-1.276534)
\psbezier[linewidth=0.024](6.54,-0.9234661)(6.14,-0.9234661)(5.74,-1.523466)(5.560209,-1.6765339)
\psbezier[linewidth=0.024](6.54,0.6765338)(6.14,0.8765338)(5.94,1.276534)(5.54,1.6765338)
\psbezier[linewidth=0.024](6.54,0.27653393)(6.34,0.27653393)(5.74,1.0765339)(5.54,1.276534)
\psline[linewidth=0.04cm,linestyle=dashed,dash=0.16cm 0.16cm](6.54,0.6765338)(8.14,0.6765338)
\psline[linewidth=0.04cm,linestyle=dashed,dash=0.16cm 0.16cm](6.54,0.27653393)(8.14,0.27653393)
\psline[linewidth=0.04cm,linestyle=dashed,dash=0.16cm 0.16cm](6.54,-0.52346605)(8.14,-0.52346605)
\psline[linewidth=0.04cm,linestyle=dashed,dash=0.16cm 0.16cm](6.54,-0.9234661)(8.14,-0.9234661)
\psbezier[linewidth=0.024](9.135333,0.82045114)(8.34,0.27653393)(8.34,-0.32346618)(9.119791,-0.876534)
\psbezier[linewidth=0.024](8.14,-0.52346605)(8.34,-0.52346605)(8.74,-0.9234661)(9.119791,-1.276534)
\psbezier[linewidth=0.024](8.14,-0.9234661)(8.54,-0.9234661)(8.94,-1.523466)(9.119791,-1.6765339)
\psbezier[linewidth=0.024](8.14,0.6765338)(8.54,0.8765338)(8.74,1.276534)(9.14,1.6765338)
\psbezier[linewidth=0.024](8.14,0.27653393)(8.34,0.27653393)(8.94,1.0765339)(9.14,1.276534)
\psbezier[linewidth=0.024](0.14466743,0.82045114)(0.94,0.27653393)(0.94,-0.32346618)(0.16020894,-0.876534)
\psbezier[linewidth=0.024](1.14,-0.52346605)(0.74,-0.7234661)(0.34,-1.123466)(0.16020894,-1.276534)
\psbezier[linewidth=0.024](1.14,-0.9234661)(0.74,-0.9234661)(0.34,-1.523466)(0.16020894,-1.6765339)
\psbezier[linewidth=0.024](1.14,0.6765338)(0.74,0.8765338)(0.54,1.276534)(0.14,1.6765338)
\psbezier[linewidth=0.024](1.14,0.27653393)(0.74,0.47653413)(0.34,1.0765339)(0.14,1.276534)
\psline[linewidth=0.04cm,linestyle=dashed,dash=0.16cm 0.16cm](1.14,0.6765338)(2.74,0.6765338)
\psline[linewidth=0.04cm,linestyle=dashed,dash=0.16cm 0.16cm](1.14,0.27653393)(2.74,-0.52346605)
\psline[linewidth=0.04cm,linestyle=dashed,dash=0.16cm 0.16cm](1.14,-0.52346605)(2.74,0.27653393)
\psline[linewidth=0.04cm,linestyle=dashed,dash=0.16cm 0.16cm](1.14,-0.9234661)(2.74,-0.9234661)
\psbezier[linewidth=0.024](3.7353327,0.82045114)(2.94,0.27653393)(2.94,-0.32346618)(3.719791,-0.876534)
\psbezier[linewidth=0.024](2.74,-0.52346605)(3.14,-0.7234661)(3.54,-1.123466)(3.719791,-1.276534)
\psbezier[linewidth=0.024](2.74,-0.9234661)(3.14,-0.9234661)(3.54,-1.523466)(3.719791,-1.6765339)
\psbezier[linewidth=0.024](2.74,0.6765338)(3.14,0.8765338)(3.34,1.276534)(3.74,1.6765338)
\psbezier[linewidth=0.024](2.74,0.27653393)(3.14,0.47653413)(3.54,1.0765339)(3.74,1.276534)
\psline[linewidth=0.05cm](0.0,-0.28346616)(9.6,-0.28346616)
\psline[linewidth=0.04cm,arrowsize=0.05291667cm 2.0,arrowlength=1.4,arrowinset=0.4]{<-}(12.115,-0.11146606)(10.515,-0.11146606)
\psbezier[linewidth=0.024](12.744667,0.82045114)(13.54,0.27653393)(13.54,-0.32346618)(12.760209,-0.87653387)
\psbezier[linewidth=0.024](13.74,0.67653394)(13.34,0.87653404)(13.14,1.276534)(12.74,1.6765339)
\psbezier[linewidth=0.024](13.74,0.27653393)(13.34,0.47653413)(12.94,1.0765339)(12.74,1.276534)
\psbezier[linewidth=0.024](13.74,-0.5234659)(13.34,-0.7234661)(12.94,-1.1234661)(12.760209,-1.2765338)
\psbezier[linewidth=0.024](13.74,-0.9234661)(13.34,-0.9234661)(12.94,-1.5234659)(12.760209,-1.6765338)
\psbezier[linewidth=0.024](16.335333,0.82045114)(15.54,0.27653393)(15.54,-0.32346618)(16.319792,-0.87653387)
\psbezier[linewidth=0.024](15.34,0.67653394)(15.74,0.87653404)(15.94,1.276534)(16.34,1.6765339)
\psbezier[linewidth=0.024](15.34,0.27653393)(15.74,0.47653413)(16.14,1.0765339)(16.34,1.276534)
\psbezier[linewidth=0.024](15.34,-0.5234659)(15.74,-0.7234661)(16.14,-1.1234661)(16.319792,-1.2765338)
\psbezier[linewidth=0.024](15.34,-0.9234661)(15.74,-0.9234661)(16.14,-1.5234659)(16.319792,-1.6765338)
\psbezier[linewidth=0.024](18.144669,0.82045114)(18.94,0.27653393)(18.94,-0.32346618)(18.16021,-0.87653387)
\psbezier[linewidth=0.024](19.14,0.67653394)(18.74,0.87653404)(18.54,1.276534)(18.14,1.6765339)
\psbezier[linewidth=0.024](19.14,0.27653393)(18.74,0.47653413)(18.34,1.0765339)(18.14,1.276534)
\psbezier[linewidth=0.024](19.14,-0.5234659)(18.74,-0.7234661)(18.34,-1.1234661)(18.16021,-1.2765338)
\psbezier[linewidth=0.024](19.14,-0.9234661)(18.74,-0.9234661)(18.34,-1.5234659)(18.16021,-1.6765338)
\psbezier[linewidth=0.024](21.735334,0.82045114)(20.94,0.27653393)(20.94,-0.32346618)(21.71979,-0.87653387)
\psbezier[linewidth=0.024](20.74,0.67653394)(21.14,0.87653404)(21.34,1.276534)(21.74,1.6765339)
\psbezier[linewidth=0.024](20.74,0.27653393)(21.14,0.47653413)(21.54,1.0765339)(21.74,1.276534)
\psbezier[linewidth=0.024](20.74,-0.5234659)(21.14,-0.7234661)(21.54,-1.1234661)(21.71979,-1.2765338)
\psbezier[linewidth=0.024](20.74,-0.9234661)(21.14,-0.9234661)(21.54,-1.5234659)(21.71979,-1.6765338)
\psline[linewidth=0.05cm,linestyle=dotted,dotsep=0.16cm](13.74,0.67653394)(15.34,0.67653394)
\psline[linewidth=0.05cm,linestyle=dotted,dotsep=0.16cm](13.94,-0.9234661)(15.34,-0.9234661)
\psbezier[linewidth=0.05,linestyle=dotted,dotsep=0.16cm](13.74,0.27653393)(14.295,0.048533965)(14.295,-0.31146604)(13.74,-0.5234659)
\psline[linewidth=0.05cm,linestyle=dotted,dotsep=0.16cm](19.34,-0.9234661)(20.74,-0.9234661)
\psline[linewidth=0.05cm,linestyle=dotted,dotsep=0.16cm](19.14,0.67653394)(20.74,0.67653394)
\psline[linewidth=0.05cm,linestyle=dotted,dotsep=0.16cm](19.14,0.27653393)(20.74,0.27653393)
\psline[linewidth=0.05cm,linestyle=dotted,dotsep=0.16cm](19.14,-0.52346605)(20.715,-0.511466)
\psbezier[linewidth=0.05,linestyle=dotted,dotsep=0.16cm](15.315,0.27653393)(14.76,0.048533965)(14.76,-0.31146604)(15.315,-0.5234659)
\end{pspicture} 
}
\end{center}
\caption{Covariances $C_{\sigma}$ (dashed lines) sandwiched in the inner product in the Cauchy-Schwarz inequality (thick line) are replaced by modified covariances $C'_{\sigma}$ (dotted lines).}
\label{covariances}
\end{figure}
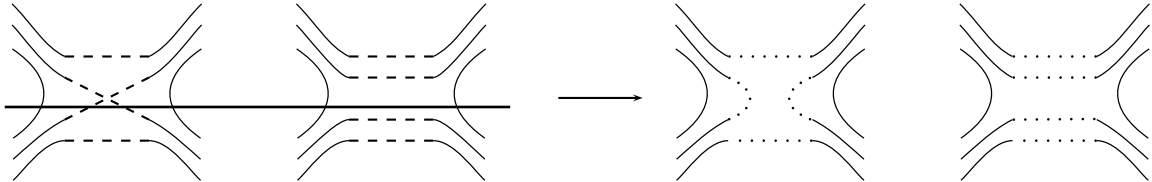

This process has two effects: thanks to the resolvent bound, it trades the original tree for two trees, with the same number of vertices, but with two resolvents replaced by the identity ($\Vert R_1\Vert\Vert R_2\Vert\leq 2$); each covariance $C_{\sigma}$ which is sandwiched in the inner product is replaced by a modified covariance $C_{\sigma}':=1+\delta\cT'$, where $\cT'$ identifies the two central arguments of $\sigma$ instead of twisting them (Fig. \ref{covariances}):
\be
T'(g_1,\dots,g_4;g'_1,\dots,g'_4):=\delta(g_{1}g_{1}'^{-1})\delta(g_{4}g_{4}'^{-1})\delta(g_{2}g_{3}^{-1})\delta(g_{2}'g_{3}'^{-1}).
\ee
% Generated with LaTeXDraw 2.0.2
% Fri Jun 26 17:36:42 CEST 2009
% \usepackage[usenames,dvipsnames]{pstricks}
% \usepackage{epsfig}
% \usepackage{pst-grad} % For gradients
% \usepackage{pst-plot} % For axes

We can iterate the process $(n-1)$ times, until all resolvents are removed, and all covariances $C_{\sigma}$ replaced by $C_{\sigma}'$. We are then left with a {\it perturbative} BFL graph, whose vertices are all $P_{2,2}^1$:

\be
\vert \cC_T(\lambda)\vert\leq 2^{n-1}\ \text{sup}_{T'\in\cT_{n}}\ \vert\cA_{T'}^{BFL}\vert,
\ee
Now, from Theorem \ref{pertBFL} we have $\text{sup}_{T'\in\cT_{n}}\ \vert\cA_{T'}^{BFL}\vert\le K^n\Lambda^{6 + 3n}$, hence $\vert \cC_T(\lambda)\vert\leq2^{n-1}K^n \Lambda^{6 + 3n}$.

Before we can conclude, we should check that this estimate also holds for the  trivial tree with just one vertex ($n=1$): since it contains no half-line, the loop vertex $\text{Log}(1+i\sqrt{\lambda}\widetilde{\Sigma})$ is not acted upon by a $\sigma$-derivative, and therefore the resolvent bound does not apply. However, using standard convexity inequalities, we have $\vert\text{Log}(1+z)\vert\le\f{1}{2}+\f{1}{2}(\vert z\vert^2+4\pi^2)$ for $z\in\mathbb{C}$, hence
\be
\Vert\text{Log}(1+i\sqrt{\lambda}\widetilde{\Sigma})\Vert\le\f{1}{2}+\f{1}{2}(\Vert\widetilde{\Sigma}\Vert^2+4\pi^2)\le\f{1}{2}+\f{1}{2}(\Vert\sigma\Vert_{\infty}^2+4\pi^2),
\ee
and thus
\be
\vert\text{Tr}\int d\nu^{(\La)}_{\delta}(\sigma)\ \text{Log}(1+i\sqrt{\lambda}\widetilde{\Sigma})\vert\le K\ \dim\cH^{(\La)}_{0}=\mathcal{O}(\La^6).
\ee

Absorbing the uniform factor $\dim\cH^{(\La)}_{0}=\mathcal{O}(\La^6)$ appearing in all estimates in the definition of a free energy per degree of freedom $\mathcal{G}_{\text{BFL}}^{(\La)}:=\f{1}{\dim\cH^{(\La)}_{0}}\mathcal{F}_{\text{BFL}}^{(\La)}$, we have proved\footnote{The Nevanlinna-Sokal criteria can be checked exactly as in the toy example.}

\begin{theorem}\label{borelBFL}
The cactus expansion of the BFL free energy per degree of freedom $\mathcal{G}_{\text{BFL}}^{(\La)}$ is uniformly convergent in a half-disk $\{\lambda\in\mathbb{C},\Re\lambda\geq0,\vert\lambda\vert\leq K\La^{-3}\}$, where it defines the Borel sum of the BFL perturbative series. %Moreover, we have the following bound on the Borel radius
%\be
%\rho^{(\La)}_{\text{BFL}}\leq K\La^{-3}.
%\ee
\end{theorem}

\section{Conclusion}

We have proved that individual graphs at order $n$ of perturbation
theory for the Boulatov and BFL models are uniformly bounded by $K^n\Lambda^{6+ 3n/2}$ and
$K^n\Lambda^{6+ 3n}$ respectively, and that the Taylor-Borel remainders for the BFL model are
bounded by $n!K^n\Lambda^{6+ 3n}$. These bounds are optimal, being saturated by a class of graphs. Since the perturbative power counting of the Boulatov model
is much better than the one of the BFL model, we conclude that the Freidel-Louapre constructive regularization with the `pillow' term
of the Boulatov model is not optimal from the perspective of power counting.

We conjecture that 

\medskip\noindent{\bf Conjecture }
Any Boulatov graph at order $n$  \emph{without two-point subgraph} (hence 
with $N\ge 4$ external legs) is bounded at fixed external legs arguments by $C^n\Lambda^{n-1}$.

Remark that this conjecture is slightly different from the conjectures of \cite{Freidel:2009hd}, as the latter paper
is more oriented towards identifying the leading graphs (called type I) and 
finding their exact power counting.

Our future program consists in

\begin{itemize}

\item Identifying a better constructive
regularization  than the BFL ``pillow term", i.e. such that the Taylor-Borel
remainders obey the same scaling in $n! C^n\Lambda^{6+3n/2 }$.

\item Proving the conjecture, presumably through a multiscale analysis, and identify a still better constructive
regularization, i.e such that the Taylor-Borel remainders of e.g. the 4-point function, 
\emph{after two point function renormalization }
 obey the same scaling in $n! C^n\Lambda^{n-1 }$.

\item Generalizing in the appropriate way these conjectures, first to the Ooguri model 
in dimension 4, then to the EPR-FK models and investigate their scaling properties
in detail through the appropriate multiscale analysis.
\resetequ\end{itemize}

\section*{Acknowledgments}

We thank Razvan Gurau, Carlo Rovelli and Simone Speziale for discussions and encouragement. This work was partially supported by the ANR (BLANL06-3 139436 LQG-2006).
\appendix
\section{Degenerate Gaussian measures}

If $A$ is a $n\times n$ positive definite matrix, a (centered, normalized) Gaussian measure $\mu^A$ over $\mathbb{R}^n$ can be defined in terms of its Radon-Nikodym derivative with respect to the Lebesgue measure $\lambda$
\be\label{Radon-Nikodym}
\f{d\mu_A}{d\lambda}(x)=\f{\sqrt{\det A}}{(2\pi)^{n/2}}e^{-\f{x.Ax}{2}}.
\ee
From this definition, it follows that the covariance matrix $C=A^{-1}$ is positive definite. Wick's theorem then implies that all higher moments are completely determined by $C$. In fact, the measure itself is uniquely characterized by the covariance, as an example of a positive solution to the Hamburger moment problem. Since the last two propositions (Wick's theorem and reconstruction from the moments) would also hold true if $C$ were positive semi-definite, a more general definition of a Gaussian measure is as follows: given a positive semi-definite matrix $C$, the Gaussian measure $\mu_C$ with covariance $C$ is the unique measure whose moments are given by Wick contractions of $C$. Integration with respect to $d\mu_C$ is then really the same
as applying the exponential of a quadratic differential operator to the integrand. More precisely, for any
smooth and summable $f$
\be\label{expquadr}  \int d\mu_C (x)\  f(x)  = 
e^{\frac{1}{2}\frac{\partial}{\partial x}  C \frac{\partial}{\partial x }  } f(x)_{\vert_{x =0}}
\ee

When $C$ has zero eigenvalues, $d\mu_{C}$ is a `degenerate Gaussian measure', with a Dirac delta component along each zero mode. In such cases, $\mu_C$ is not absolutely continuous with respect to the Lebesgue measure, and (\ref{Radon-Nikodym}) does not make sense, while (\ref{expquadr}) does.

A first useful example of such a situation is the {\it replica trick}
considered in section \ref{constructive}, which is the key to the rigorous construction
of connected functions in QFT. 
It relies on the existence of the degenerate Gaussian measure
defined by the rank-$1$ matrix $C_{ij}=1$.  Its eigenvalues are 0, with multiplicity $n-1$, and $n$, with multiplicity $1$ and eigenvector $(1,\dots,1)$. It follows that for any smooth and summable function $f$ in
$n$ variables:
\be
\int_{\mathbb{R}^n} d\mu_C(x)\ f(x)=\f{1}{\sqrt{2\pi}}\int_{\mathbb{R}}d\sigma\ e^{-\sigma^2/2}f(\sigma,\dots,\sigma).
\ee
The interest of such a seemingly trivial measure is the possibility to perturb {\it off diagonal}
coefficients of this Gaussian measure to break the replica symmetry.\footnote{A. Connes always remarks, especially when discussing Galois's work, that it is not
symmetry which is important, but how to break it.}

Another instructive example is precisely given by group field theory.
It is often formulated as a functional integral over 
invariant fields, that is loosely speaking as
\be  \label{loosegft}
\int D_{inv}\phi\  e^{- \phi^2/2 - \lambda \phi^{d+1}},
\ee
where the tensorial nature of the $\phi^{d+1}$ term reflects the gluing of
$d$-simplices along common $(d-1)$-faces, and the Gaussian part looks like
a trivial ``local" mass term. The nontrivial content of the theory is then completely hidden
in the ``Lebesgue measure" $D_{inv} (\phi)$, which restricts functional integration
to fields invariant under e.g. diagonal right multiplication:
\be \label{invconstraints}
 \phi(g_1, ..... g_d) = \phi (g_1h, ... g_dh) \quad \forall h \; .
\ee

But this loose writing is confusing. Remember that
in the case of an ordinary field theory, such as the
ordinary $\phi^4$ theory, the loose writing of functional integration as
\be
\int D\phi\  e^{- \int \f{1}{2}\phi  (-\Delta + m^2) \phi  - \lambda \int \phi^{4}} 
\ee
is already misleading because there is no well-defined ``infinite dimensional
Lebesgue measure" $D\phi = \prod_x d\phi(x)$ in any reasonable sense.
As well known in constructive theory,   $\int D\phi$ and  $ e^{- \f{1}{2}\int \phi  (-\Delta + m^2) \phi}$
should be combined  into the Gaussian measure $d\mu_C$ with propagator
$C = (-\Delta + m^2)^{-1}$ which is 
mathematically well-defined as a measure on a space of distributions through the Minlos Theorem. One should therefore write
\be \label{betterqft} \int d\mu_C(\phi)\  e^{- \lambda \int\phi^{4}}.
\ee

Then one identifies the problems:  the interaction term in dimension two or more
 is not Radon-Nikodym with respect to that Gaussian measure because of ultraviolet divergences, and moreover a thermodynamic limit has also to be performed because the infinite volume
integral $\int_{\mathbb{R}^d}\phi^{4}$ diverges with probability one. The expression (\ref{betterqft}) at least opens a sensible avenue, which is to apply an ultraviolet cutoff 
on $C$ in $d\mu_C$, start with a finite volume instead of ${\mathbb R}^d$,
and perform the two limits in the right
way (that is step by step, according to the renormalization group).

In (\ref{loosegft}), $D_{inv}\phi $ is the product of the previously ill-defined
``infinite dimensional Lebesgue measure" $D \phi$ by $\delta$-functions constraints implementing
the constraints (\ref{invconstraints}). So in a way it has no Radon-Nikodym density, but with
respect to a measure which is not well defined! This is why the writing (\ref{loosegft})
is doubly confusing. The way out is again to remark that because $\delta$-functions are
Gaussian measure, one can combine the constraints, the `mass' term and the 
Lebesgue measure into a Gaussian measure $d\mu_C$
with propagator
\be  C(g_1,..., g_d; g'_1,..., g'_d) = \int dh\ \delta (g_1^{-1} h g'_1)\cdots \delta (g_d^{-1} h g'_d)
\ee
which is now well-defined.

\section{Trees and forests}
%Mathematicians 
%usually prefer {\it oriented} graphs without tadpoles since they can
%be neatly characterized through their incidence
%matrix ${\epsilon_{v e}}$.  It is the rectangular $E$ by $V$ matrix 
%with indices running over vertices and lines respectively, such that 
%\begin{itemize}
%\item
%${\epsilon_{ve}}$ is +1 
%if $e$  starts at $v$, 
%\item
%${\epsilon_{ve}}$ is -1 if $e$ ends at $v$,
%\item
%${\epsilon_{ve}}$ is  0 otherwise\footnote{
%The incidence matrix for a tadpole $e$ at vertex $v$ can be 
%defined eg with $\epsilon_{ve} =2$, $\epsilon{v'e} =0$ for $v' \ne v$.}.
%\end{itemize}

%An unoriented graph has incidence matrix $\eta_{ve} = \vert \epsilon_{ve}\vert$.
%The coordination (degree) of a vertex is the number of lines which hook at it,
%hence $c_v= \sum_v \eta_{ve}$. The {\it complete graph} over $n$ vertices is the unique graph 
%with $n(n-1)/2$ lines in which any pair of vertices is joined by a line. 

%A line whose removal increases (by one) the number of connected parts 
%of the graph is called a one-particle-reducible line (bridge).
For convenience, we collect here some basic definitions of graph theory. We use the quantum field theorists' vocabulary, recalling the mathematicians' in parentheses.

A {\it graph} (pseudograph) $G$ is a set of vertices $V$ and of lines (edges) $E$, together with 
an incidence relation between them, possibly with 
several lines connecting the same vertices (multiple edges) and lines connecting a vertex to itself -- tadpoles (loops). A {\it subgraph} $G'$ of $G$ is a subset of edges of $G$, together with the attached vertices. A {\it loop} (cycle) is a connected subset of $n$ lines and $n$ vertices 
which cannot be disconnected by removing any line. 

A {\it forest} is a graph without loops. A \emph{spanning forest} of $G$ is a sub-forest of $G$ 
that contains all the vertices of $G$. A {\it tree} is a connected forest, or equivalently a graph with $\vert E\vert=\vert V\vert-1$. A vertex with coordination $1$ in a tree is a {\it leaf}. A {\it rooted tree} is a tree with a distinguished vertex, its {\it root}.

A useful result in graph theory is Cayley's formula, giving the number $T_n$ of different labeled trees on $n$ vertices,
\be\label{cayley2}
T_n=n^{n-2}, 
\ee
and the number of such trees with fixed coordinations $k_v$
\be\label{cayley}
T_n(\{k_v\})=\f{n!}{\prod_{v=1}^n(k_v-1)!}.
\ee

Four proofs of this formula are given in \cite{TheBook}. %A simple one is Pitman's, which counts in two different ways the number of different sequences of directed lines that can be added to an empty graph on $n$ vertices to form from it a rooted tree. One way to form such a sequence is to start with one of the $T_n$ possible unrooted trees, choose one of its $n$ vertices as root, and choose one of the $(n - 1)!$ possible sequences in which to add its $n - 1$ lines. Therefore, the total number of sequences that can be formed in this way is $T_nn(n - 1)! = T_nn!$. Another way to count these line sequences is to consider adding the edges one by one to an empty graph, and to count the number of choices available at each step. If one has added a collection of $n - k$ lines already, so that the graph formed by these lines is a rooted forest with $k$ trees, there are $n(k - 1)$ choices for the next edge to add: its starting vertex can be any one of the $n$ vertices of the graph, and its ending vertex can be any one of the $k$ roots other than the root of the tree containing the starting vertex. Therefore, if one multiplies together the number of choices from the first step, the second step, etc., the total number of choices is
%\be
%\prod_{k=2}^nn(k-1)=n^{n-1}(n-1)!=n^{n-2}n!.
%\ee
%Equating these two formulas for the number of edge sequences results in Cayley's formula.

\section{Borel resummation}

It is a classic theorem of Borel that there is an infinite number of smooth real functions 
asymptotic to {\it any} power series. 
But analytic functions are {\it rigid}: within their domain of analyticity, all the information about the function
is encapsulated in the countable list of its Taylor coefficients.
Ordinary summation provides a 
one-to-one correspondence (at least inside a convergence disk) between 
a convergent power series and a unique ``preferred" function 
asymptotic to that series, namely the analytic one.

Borel summability is a natural way to extend this one-to-one 
correspondence between convergent series and analytic functions. When an 
analytic function admits a Taylor series around a point on the 
$boundary$ of its domain of analyticity, this series has zero radius of convergence.
But under some conditions, all the information about the function 
can be still be encapsulated by the series. Borel summation picks up the unique ``preferred" function 
asymptotic to that series, namely its Borel sum. 

Borel summability is ubiquitous in theoretical physics, where most expansions have zero radius
of convergence. That it was the case of the perturbative series of quantum field theory was realized by Dyson back in 1952 \cite{PhysRev.85.631}.

The most natural criteria for Borel summability were formulated by Nevanlinna \cite{Nevanlinna}, and rediscovered by Sokal \cite{Sokal:1980ey}:

\begin{theorem}[Nevanlinna-Sokal, direct] 

\noindent Let $f$ be analytic in the disk 
$C_R := \{ y \vert {\rm Re}\, y^{-1} > 1/R\}$. 
Suppose $f$ admits an asymptotic power series $\sum a_k y^k  $
(its Taylor series at the origin)
\be  \label{bor1}
f(y) = \sum_{k=0}^{r-1} a_k y^k + R_r(y)   
\ee
such that the bound
\be  \label{bor2}
\vert R_r (y) \vert \le c \sigma^r r! \vert y \vert^r  
\ee
holds uniformly in $r$ and $y \in C_R$, for some constants $\sigma$ and $C$.
Then $f$ is Borel summable, {\it i.e.} the power series $B(t):=\sum_k a_k {
t^k \over k!}$ converges for $\vert t \vert < {1 \over \sigma}$, and admits an analytic continuation in the strip 
$S_{\sigma} := \{t \vert {\rm \ dist \ } (t, {\mathbb R}^+) < {1 \over \sigma}\}$, 
satisfying the bound
\be \label{bor3}
\vert B(t)  \vert \le { \rm const.} e^{t \over R} \quad {\rm for \ } t \in {{\mathbb R}}^+  
\ee
Moreover, $f$ is represented in $C_R$ by the absolutely convergent integral
\be  \label{bor4}
f(y) = {1 \over y} \int_{0}^{\infty}dt\ e^{-{t \over y}}  B(t)
\ee
\end{theorem}
$B$ is then called 
the Borel transform of $f$, and the complex $t$ plane is called the Borel plane. 

There is 
a reciprocal to this theorem:

\begin{theorem}[Nevanlinna-Sokal, reciprocal] 
Consider the power series $\sum a_k y^k$. If the power series $ \sum a_k {t^k \over k!}$ converges in a disk $\vert t \vert < {1 \over \sigma}$, admits an analytic continuation 
$B(t)$ in the strip $S_{\sigma}$ and satisfies the bound (\ref{bor3}) in this 
strip, then the function $f$ defined by the integral representation (\ref{bor4}) is analytic in 
$C_R$, has $\sum a_k y^k$ as Taylor series at the origin and satisfies the 
uniform remainder estimates. 
\end{theorem}

In this case we say  that the series $\sum a_k y^k$ is Borel summable, and call the series $\sum a_k 
{t^k \over k!}$ its Borel transform and that the function $f$ its Borel 
sum. 

In conclusion, Borel summable series and Borel summable functions are in 
correspondence just like are ordinary series and germs of analytic functions.
However, the analytic continuation in the Borel strip involved in the 
construction of the function from its series is usually intractable, and so in practice only the direct theorem is used. 

For interactions of higher degree than $\phi^4$, which will be required for group field theory in dimension 4 and higher, extended notions of Borel resummation, such as Borel-LeRoy
resummation of order $\alpha$, must be applied. In essence, they combine suitable conformal transformations of the standard disk $C_R$ with the standard Nevanlinna-Sokal criteria.

\end{document}